%% file: membranesquivers11.tex
\documentclass[11pt,a4paper,english]{article}
\usepackage{graphicx}
\usepackage{amssymb,latexsym}
\usepackage{amsmath}
\usepackage{epsf}
\usepackage{pdfsync}
\usepackage{epsfig}
\usepackage[parfill]{parskip}
\usepackage[matrix,arrow,color]{xy}
\usepackage[usenames]{color}
\usepackage{mathrsfs}
\usepackage[mathcal]{euscript}
\usepackage[font={small}]{caption}
\usepackage{subcaption}
\usepackage[export]{adjustbox}
\usepackage{float}
\usepackage{wrapfig}
\usepackage{microtype}
\usepackage{slashed}
\usepackage [latin1]{inputenc}
\usepackage[bookmarks=false]{hyperref}

\input{xy}
\xyoption{all}

\input{epsf}

\makeatletter

\usepackage{setspace}

\usepackage{lscape}

\catcode`@=11
\renewcommand{\section}
{\@startsection{section}{1}{0pt}{\medskipamount}{\medskipamount}{\large\bf}}
\makeatletter\renewcommand{\subsection}
{\@startsection{subsection}{2}{\z@}{-3.25ex plus -1ex minus -.2ex}
{1.5ex plus .2ex}{\it }}

\numberwithin{equation}{section}
\catcode`@=12

\newcommand{\ban}{\begin{eqnarray}}
\newcommand{\ean}{\end{eqnarray}}

\newcommand{\IC}{\mathbb{C}}

\newcommand{\cI}{{\cal I}}
\newcommand{\cW}{{\cal W}}
\newcommand{\cN}{{\cal N}}
\newcommand{\cM}{{\cal M}}

\newcommand{\cE}{{\cal E}}
\newcommand{\cO}{{\cal O}}

\newcommand{\cQ}{{\cal Q}}

\newcommand{\cF}{{\cal F}}

\newcommand{\cK}{{\cal K}}

\newcommand{\cV}{{\cal V}}

\newcommand{\sfA}{{\mathsf{A}}}
\newcommand{\sfe}{{\mathsf{e}}}
\newcommand{\sfa}{{\mathsf{a}}}

\newcommand{\sfP}{{\mathsf{P}}}
\newcommand{\sfZ}{{\mathsf{Z}}}
\newcommand{\sfQ}{{\mathsf{Q}}}
\newcommand{\sfD}{{\mathsf{D}}}
\newcommand{\sfS}{{\mathsf{S}}}

\newcommand{\sfW}{{\mathsf{W}}}

\newcommand{\scrM}{{\mathscr{M}}}

\newcommand{\DT}{{\tt DT}}

\newcommand{\Hecke}{\mathcal{H\kern-.77em H}}

\newcommand{\ahat}[1]{{  \frac{\left( \frac{ \kappa}{#1} \right)^{1/2} - \left( \frac{#1}{ \kappa}\right)^{1/2} }{\left( #1 \right)^{1/2} - \left( #1 \right)^{-1/2}}}}

\newcommand{\complex}{{\mathbb C}} 
\newcommand{\zed}{{\mathbb Z}} 
\newcommand{\real}{{\mathbb R}} 
\newcommand{\torus}{{\mathbb T}}

\def\e{{\,\rm e}\,}

\def\ii{{\,{\rm i}\,}}
\def\dd{{\rm d}}

\newcommand{\Hom}{\mathrm{Hom}}

\newcommand{\End}{\mathrm{End}}
\newcommand{\sign}{\mathrm{sgn}}

\newcommand{\Ext}{\mathrm{Ext}}

\def\beq{\begin{equation}}
\def\bee{\begin{equation}}
\def\eeq{\end{equation}}
\def\bea{\begin{eqnarray}}
\def\eea{\end{eqnarray}}
\def\bd{\begin{displaymath}}
\def\ed{\end{displaymath}}

\newcommand{\Cint}{\int\kern-10.5pt-\kern7pt}

\newcommand{\PP}{{\mathbb{P}}}

\newcommand{\be}{\begin{equation}}
\newcommand{\ee}{\end{equation}}
\newcommand{\bal}{\begin{align}}
\newcommand{\eal}{\end{align}}

\newcommand\fverbit{\egroup\item[\fbox{\unhbox\pippobox}]}
\newbox\pippobox

{

\def\be{\begin{equation}}
\def\ee{\end{equation}}
\def\bea{\begin{eqnarray}}
\def\eea{\end{eqnarray}}

\newtheorem{conjecture}{Conjecture}

\begin{document}

\begin{titlepage}
\setcounter{page}{1}

\vskip 5cm

\begin{center}

\vspace*{3cm}

{\Huge On the M2--Brane Index on \\[15pt] Noncommutative Crepant Resolutions}

\vspace{15mm}

{\large\bf Michele Cirafici}
\\[6mm]
\noindent{\em Dipartimento di Matematica e Geoscienze, Universit\`a di Trieste, \\ Via A. Valerio 12/1, I-34127 Trieste, Italy, 
\\ Institute for Geometry and Physics (IGAP), via Beirut 2/1, 34151, Trieste, Italy
\\ INFN, Sezione di Trieste, Trieste, Italy 
}\\[4pt] Email: \ {\tt michelecirafici@gmail.com}

\vspace{15mm}

\begin{abstract}
\noindent

On certain M-theory backgrounds which are a circle fibration over a smooth Calabi-Yau the quantum theory of M2 branes can be studied in terms of the K-theoretic Donaldson-Thomas theory on the threefold. We extend this relation to noncommutative crepant resolutions. In this case the threefold develops a singularity and classical smooth geometry is replaced by the algebra of paths of a certain quiver. K-theoretic quantities on the quiver representation moduli space can be computed via toric localization and result in certain rational functions of the toric parameters. We discuss in particular the case of the conifold and certain orbifold singularities.

\end{abstract}

\vspace{15mm}

\today

\end{center}
\end{titlepage}

\tableofcontents

\section{Introduction}

One of the most interesting aspects of string/M-theory is that it is capable of describing quantum degrees of freedom in spacetimes where the manifold structure breaks down. For example in the moduli spaces of vacua of Calabi-Yau compactifications one can generically find points where the manifold develops a singularity. The local dynamics near such a point is often captured by a non-compact threefold. For example the quintic contains a conifold point, where a rigid rational curve shrinks to zero size. The local dynamics can be studied using strings on the resolved conifold, by sending to zero the K\"ahler volume of the curve. 

The situation is less clear in M-theory where we lack a complete description of the fundamental degrees of freedom. In \cite{Nekrasov:2004vv,NO} it was proposed that certain aspects of the theory, namely the supersymmetric index of the membrane, can be computed explicitly using K-theoretic Donaldson-Thomas theory. In their framework, one does not need the 3-fold to be Calabi-Yau, but only that the ambient 5-fold is Calabi-Yau, as long as the overall geometry is smooth. Computing the index of the membrane is equivalent to enumerating supersymmetry protected states with certain quantum numbers and therefore provides a window on the fundamental degrees of freedom and their supersymmetric excitations.

In this note we study K-theoretic Donaldson-Thomas theory defined on noncommutative crepant resolutions of Calabi-Yau singularities as a model for the M2-brane index in M-theory. Recent progress in K-theoretic countings of BPS states have been made in the context of smooth geometries and fourfolds \cite{Benini:2018hjy,Bonelli:2020gku,DelZotto:2021gzy,Descombes:2021snc,Fasola:2020hqa, Kononov:2019fni,Mozgovoy:2020has,Nekrasov:2017cih,Nekrasov:2018xsb,Pomoni:2021hkn} as well as defects in quantum field theory \cite{Cirafici:2019otj}.

In many cases when a toric threefold develops a singularity the local geometry is replaced by algebraic quantities. This is the case for the conifold or for resolved orbifolds, when the orbifold group is subgroup of $\mathrm{SU(3)}$. In these examples there exists a noncommutative algebra $\sfA$ which contains the local singularity as its center. The algebra itself plays the role of a resolution of the singularity and it is called a noncommutative crepant resolution \cite{vandenbergh}. D-branes can be consistently studied in such backgrounds. Indeed the derived category of modules over $\sfA$ is derived equivalent to the derived category of coherent sheaves over the smooth threefold. In other words such backgrounds are required for the consistency of the compactification.

The algebra $\sfA$ can be seen as the Jacobian path algebra of a quiver with relations. The low energy dynamics of BPS states can be described in terms of a certain quantum mechanical model associated with this quiver. When one restricts attention to BPS states formed by a gas of D2/D0 branes bounded to a single D6 brane the computation can be carried out exactly. The presence of the D6 brane modifies the quiver via a framing node. The toric symmetry of the original threefold descends to the framed quantum mechanics model and its partition function, the index of D-brane states, can be computed directly via toric localization in many cases \cite{szendroi,Cirafici:2010bd}.

For toric threefolds the enumeration of BPS bound states in the type II string at large radius and on noncommutative crepant resolutions are related by wall crossing \cite{Jafferis:2008uf,Nagao:2010kx}. Consistency requires that a similar relation should hold in K-theoretic enumerative problems. The large radius index for membranes can be computed within the formalism of \cite{NO}. In this proposal the enumeration of supersymmetric membrane states is again given in terms of a melting crystal, where each crystal configuration is weighted by a certain function. The result depends explicitly on the toric parameters of the large radius threefold.

We expect that a similar computation should exist on M-theory backgrounds where the toric threefold develops a singularity and the geometric setup is replaced by the algebra $\sfA$. In this note we develop a formalism to carry out these computations and conjecture explicit closed forms for the membrane partition function, as plethystic exponentials of rational functions. We postpone the analysis of the wall crossing behaviour of the membrane index to a follow up paper \cite{MWC}.

This note is organized as follows. In Section 2 we review BPS states on local threefolds and the M2-brane index. Section 3 contains some background material on noncommutative crepant resolutions. In Section 4 and 5 we compute the M2-brane index for the noncommutative conifold and orbifold singularities respectively. In Section 6 we conclude with a discussion and some open problems.

To make the paper easier to read the computations are detailed explicitly in the supporting \textsc{mathematica} files \cite{mathe}.

\section{The M2--brane index for local threefolds}

In this section we briefly review BPS states on local threefolds and the M2-brane index construction of \cite{NO}. BPS bound states for toric Calabi-Yau are captured by the localization of a certain gauge theory living on the D6 brane worldvolume \cite{Iqbal:2003ds,Ooguri:2009ijd}. In particular for bound states with a single D6 brane, this is Donaldson-Thomas theory of ideal sheaves. The proposal of \cite{NO} is that the K-theoretic version of this story captures the membrane contribution to the index of M-theory.

\subsection{BPS states in type IIA}

A concrete way to set up BPS counting problems on toric threefolds is to consider the type IIA string on a compact Calabi-Yau and then take a local limit. After the first step the resulting low energy theory is $\cN=2$ supergravity and its BPS states can be engineered by wrapping D-branes on holomorphic cycles of the Calabi-Yau geometry. The theory has a moduli space of vacua which parametrizes complexified K\"ahler deformations of the Calabi-Yau manifold. Which BPS states are physically realized as stable states depends on the K\"ahler moduli. The K\"ahler moduli space has an intricate chamber structure. As the K\"ahler moduli are varied one will generically cross walls of marginal stability and the physical BPS spectrum will change according to the wall-crossing formula for generalized Donaldson-Thomas invariants \cite{KS}.

Ordinary Donaldson-Thomas theory concerns the large radius chamber, where supersymmetric BPS states are bound states of D6-D4-D2-D0 branes labelled by a charge vector $\gamma \in \Gamma = H^{\rm even} (X , \zed)$ associated via Poincar\'e duality to holomorphic cycles in $H_{\rm even} (X , \zed)$. Choosing an electro-magnetic splitting one can write
\be
\gamma = p^0 + P^A \, D_A + Q_A \, \tilde{D}^A + q_0 \dd V
\ee
where $\{ D_A \}$ is a basis of $H^2 (X ; \zed)$, $\{ \tilde{D}^A \}$ the dual basis of $H^4 (X ; \zed)$ and $\dd V$ the volume element. The lattice of charges is endowed with a symplectic product
\be
\big\langle\gamma\,,\,\gamma'\,\big\rangle_{\Gamma} = \int_X\, \gamma\wedge (-1)^{{\rm
    deg}/2}\, \gamma' \ .
\label{DSZint}
\ee
In the large radius limit one can neglect worldsheet instanton corrections and the holomorphic central charge is given by
\begin{equation}
Z_X (\gamma;t) = -\int_{X}\, \gamma \wedge \e^{-t}
\end{equation}
where $t = B+ \ii J$ is the complexified K\"ahler modulus consisting of the
background supergravity two-form
$B$-field and the K\"ahler $(1,1)$-form $J$ of~$X$. Physically the theory can be understood from the point of view of the D6-branes as a $U(n)$ supersymmetric Yang-Mills theory which localizes onto solutions of the Donaldson-Uhlenbeck-Yau equations. 

At this point of the moduli space the theory is well defined in full generality for an arbitrary charge vector and a compact threefold \cite{thomas}. For toric threefolds and a particular choice of the charge vector, a single D6 brane and no D4 branes, the gauge theory becomes abelian and the enumerative problem of Donaldson-Thomas invariants is equivalent to the enumerative problem of Gromov-Witten invariants, and therefore to the topological string \cite{Maulik:2003rzb,Maulik:2008nu}. As the moduli are varied we will cross walls of marginal stability and the enumerative problem will change accordingly. 

To properly define non-compact D6 branes one needs to take the limit from a compact Calabi-Yau; then this construction results in a certain extension of the local K\"ahler moduli space \cite{Jafferis:2008uf}.
%
%
%
By moving through a path in this space of parameters we will generically cross many walls of marginal stability producing new BPS state counting problems. In this paper we will only focus on one such chamber, the noncommutative crepant resolution chamber. A more detailed study of wall-crossing properties will appear elsewhere \cite{MWC}.

\subsection{The M2--brane index at large radius} \label{M2braneindex}

The above discussion can be lifted to M-theory by considering a background $\tilde{Z}$ which is a $S^1$ bundle over a Calabi-Yau 5-fold $Z$. To begin with, one can pick $Z$ to be $X \times \complex^2$, with $X$ a smooth toric Calabi-Yau threefold.  Then such an M-theory background can be explicitly constructed as the total space of this fibration and it includes a factor with a Taub-NUT metric, whose $U(1)$ isometry is identified with the M-theory circle $S^1$. The $S^1$ fibration induces a $\complex^\times_q$ action which acts with weights $q$ and $q^{-1}$ on the $\complex^2$ factor and leaves $X \subset Z$ fixed. 

%

The fundamental insight of \cite{NO} is that the M2 brane index for M-theory on $\tilde{Z}$ can be recast in terms of the K-theoretic Donaldson-Thomas theory of $X$. We will now highlight the main points of the construction following \cite{Okounkov}. K-theoretic Donaldson-Thomas theory can be understood as a generalization of the ordinary moduli problem of BPS states counting. Roughly speaking this is usually formulated in terms of a moduli space $M$ obtained as the locus $s^{-1} (0) \subset \widetilde{M}$ where $s$ is a section of $\cE$, the obstruction bundle. The structure sheaf $\cO_M$ is the 0-th cohomology of the Koszul complex 
\be
0 \rightarrow \bigwedge^{rk} \cE^\vee \rightarrow \cdots \rightarrow \bigwedge^2 \cE^\vee \rightarrow \cE^\vee \rightarrow \cO_{\widetilde{M}} \rightarrow 0 \, .
\ee
For the purpose of virtual countings it is better to introduce the virtual sheaf $\cO_M^{\rm vir}$ as \cite{fantechi}
\be
\cO_M^{\rm vir} = [  \bigwedge^{rk} \cE^\vee \rightarrow \cdots \rightarrow \bigwedge^2 \cE^\vee \rightarrow \cE^\vee \rightarrow \cO_{\widetilde{M}} 
 ] \, .
\ee

However in physics one typically needs a complex constructed out of a Dirac operator. The main difference is that in this setup the relevant object is not $\cO_M^{\rm vir} $ but the symmetrized structure sheaf $\widehat{\cO}_M^{\rm vir} = (-1)^m \cK_{\rm vir}^{1/2} \otimes \cO_M^{\rm vir}$. Here $(-1)^m$ is an appropriate sign, which depends on the details of the moduli problem \cite{NO}, while $\cK_{\rm vir}^{1/2}$ is the square root of the virtual canonical bundle.


To be more concrete we will now specialize to the case $X= \complex^3[z_1,z_2,z_3]$. This space carries a natural action by the torus $\mathbb{T} = \mathrm{diag} (t_1 , t_2 , t_3) \subset \mathrm{GL} (3)$ with determinant $\kappa$, given by $z_i \longrightarrow t_i \, z_i$ with $i=1,2,3$. In this case the relevant moduli space is the Hilbert scheme of points $M = \mathrm{Hilb} (\complex^3 , n)$. We can see $\mathrm{Hilb} (\complex^3 , n)$ as the locus in $\widetilde{M} = \mathrm{Hilb} (\mathrm{free}_3 , n)$ (the space of $n \times n$ matrices $B_1 , B_2  , B_3 \in \End (\complex^n)$ together with a cyclic vector $v \in \complex^n$ and modulo the action of $GL(n)$) cut out by the critical locus $s = \partial \, \sfW = 0$, with
\be
\sfW = \left( B_1 \, B_2 \, B_3 - B_1 \, B_3 \, B_2 \right) \, .
\ee 
The toric action on $X$ lifts to $\mathrm{Hilb} (\complex^3 , n)$ as $B_i \longrightarrow B_i \, t_i$ for $i=1,2,3$. 

Now the obstruction bundle $\cE$ is the twisted cotangent bundle of $\widetilde{M}$, $\kappa \otimes T^* \widetilde{M}$. In this case the sheaf $\cK_{\rm vir}$ is the determinant of the dual of the virtual tangent space $T^{\rm vir}_M = \left( T_{\widetilde{M}} - \kappa \otimes T^*_{\widetilde{M}} \right) \vert_{M}$. In more general situations the sheaf $\widehat{\cO}_M^{\rm vir}$ can be generalized and the relevant object to consider is its tensor product  with a tautological bundle \cite{NO}.

The K-theoretic Donaldson-Thomas partition function is then defined as
\be
\sfZ_{\complex^3} (q , \{ t_i \}) := \chi \left( \mathbb{M} , \widehat{\cO}^{\rm vir} \right) = \sum_n (- q)^n \chi \left(  \mathrm{Hilb} (\complex^3 , n) , \widehat{\cO}^{\rm vir} \right)
\ee
where $\mathbb{M} = \bigsqcup_n \, \mathrm{Hilb} (\complex^3 , n)$. The counting parameter $q$ can be introduced by a minor modification of the symmetrized virtual structure sheaf.

This partition function can be computed directly via virtual localization as follows. The information on deformations and obstructions of the moduli space is encoded in the virtual tangent space
\be \label{Twdeco}
T^{\rm vir}_\pi =  Def - Obs =  \sum_i a_i - \sum_i b_i = \sum_i w_i - \sum_i \kappa/w_i
\ee
where $a_i$ and $b_i$ are generic placeholders for the toric weights of the deformations and the obstructions. In the case at hand these have the form $w_i$ and $\kappa/w_i$, with $w_i$ monomials in the toric weights. In localized K-theory we can write
\be
\cO^{\rm vir}_{\mathrm{Hilb} (\complex^3 , n)} = \sum_{|\pi| = n} \, \cO_{I_\pi} \, \prod_i \frac{1 - \frac{w_i}{\kappa}}{1 - w_i^{-1}}
\ee
where fixed points in $\mathrm{Hilb} (\complex^3 , n)$ are labelled by monomials ideals $I_{\pi}$ associated with three dimensional partitions $\pi$ with a fixed number of boxes $|\pi| = n$.
 
The symmetrized virtual structure sheaf is now $\widehat{\cO}^{\rm vir} = \cO^{\rm vir} \otimes (\cK^{\rm vir})^{1/2}$ where
\be
\cK^{\rm vir} = \frac{\det Obs}{\det Def} = \frac{\det (T^* \widetilde{M} \otimes \kappa) \vert_{M}}{\det T \widetilde{M} \vert_{M}} \, .
\ee

In the case at hand for $M = \mathrm{Hilb} (\complex^3 , n)$
\be
\left( T_{\widetilde{{M}}} \big\vert_{{M}} \right)_{\pi}= (\complex^3-1) \otimes \overline{V} \otimes V + V \, ,
\ee
as a local character, and therefore
\be \label{TvirC3}
T^{\rm vir}_\pi = \left( T_{\widetilde{{M}}} \big\vert_{{M}} - \kappa \otimes T^*_{\widetilde{{M}}} \big\vert_{{M}} \right)_{\pi} = V - t_1 t_2 t_3 \, \overline{V} - V \otimes \overline{V} \, (1-t_1) (1-t_2) (1-t_3) \, ,
\ee
where
\be
V = \sum_{(i,j,k) \in \pi} t_1^{-i} \, t_2^{-j} \, t_3^{-k}
\ee
is the character of the subscheme cut-out by the $\torus$-fixed ideal sheaf $\cI_\pi$. 
Here we have introduced the involution $\overline{t}_i = t_i^{-1}$ so that 
\be \label{Vbar}
\overline{V} = \sum_{(i,j,k) \in \pi} t_1^{i} \, t_2^{j} \, t_3^{k} \, .
\ee

Here,
\be
\left( \mathcal{K}_{\rm vir}^{1/2} \right)_{\pi} = \mathrm{det}^{-1/2} \, T^{\rm vir}_\pi = \frac{\det^{1/2} Obs}{\det^{1/2} Def} = \frac{\prod_i \sqrt{\frac{\kappa}{w_i}}}{\prod_i \sqrt{w_i}} \, .
\ee
The net effect of the twist by $\cK^{1/2}$ is to replace the contribution of each fixed point by
\be
\frac{1-w/\kappa}{1-w^{-1}} \longrightarrow \frac{(k/w)^{1/2} - (w/k)^{1/2}}{w^{1/2}- w^{-1/2}} = \widehat{\mathsf{a}} (w - \kappa/w) \, .
\ee
We have introduced the localized A-roof genus 
\be \label{aroof}
\widehat{\mathsf{a}} (w) = \frac{1}{w^{1/2} - w^{-1/2}}
\ee
which enjoys the properties $\widehat{\mathsf{a}} (w_1 + w_2) = \widehat{\mathsf{a}} (w_1) \, \widehat{\mathsf{a}} (w_2)$ and in particular $\widehat{\mathsf{a}} (2 w) = \widehat{\mathsf{a}} (w)^2$.

At a $\torus$-fixed point the virtual tangent space decomposition can be written schematically as in \eqref{Twdeco}; then by using the A-roof genus   \eqref{aroof} we can write
\be
\hat{\mathsf{a}} (T_\pi^{\rm vir}) = \prod_i \frac{(\kappa / w_i)^{1/2} - (w_i / \kappa)^{1/2}}{w_i^{1/2} - w_i^{-1/2}}
\ee
so that
\be
\widehat{\cO}^{\rm vir}_{\mathrm{Hilb} (\complex^3 , n)} =  \sum_{|\pi| = n} (-1) ^n \, \cO_{I_\pi} \, \hat{\mathsf{a}} (T_\pi^{\rm vir})
\ee
in localized K-theory. Following \cite{NO} the full result can be written in closed form in terms of the Nekrasov function
\be
\cF_{\complex^3} [ t_1,t_2,t_3,t_4,t_5] := \frac{\prod_{i < j \le 3} (t_i t_j)^{1/2}- (t_i t_j)^{-1/2}}{\prod_i^5 (t^{1/2}_i - t_i^{-1/2})}
\ee
as
\be
\sfZ_{\complex^3} (q ; \{ t_i \}) = \sum_n  q^n \chi (\mathrm{Hilb} \left( \complex^3 , n \right), \widehat{\cO}^{\rm vir}) = \sfS^\bullet \cF_{\complex^3} [ t_1,t_2,t_3,t_4,t_5] 
\ee
and gives the generating function of the degree 0 K-theoretic Donaldson-Thomas invariants of $\complex^3$. In the right hand side the toric weights $t_4$ and $t_5$ act on the $\complex^2$ side of the fivefold $Z = \complex^5$. They are related to the left hand side parameters by $t_4 = q / \sqrt{\kappa}$ and $t_5 = 1/(q \sqrt{\kappa})$, with $\kappa = t_1 t_2 t_3$.

The symmetrization operation $\sfS^\bullet$, or plethystic exponential, is defined as follows. Let $\cF [t_1 , \cdots , t_k ]$ a function of $k$ arguments; then we set
\be
\sfS^\bullet \, \cF [t_1 , \cdots , t_k ] := \exp \left( \sum_{n=1}^\infty \,  \frac{1}{n} \cF [t_1^n , \cdots , t_k^n ]\right) \, .
\ee

%

A generating function for refined Donaldson-Thomas invariants can be obtained from the full K-theoretic result \cite{NO}. To this end one picks a subtorus of $\torus$ defined by sending the toric weights ${t_1 ,t_2,t_3}$ to infinity or to zero in such a way that their product $\kappa$ is kept constant. The result will in principle depend on the choice of such a torus.

Before explaining this let us quickly review the various refinements of the MacMahon function. One can define a one parameter family of refined MacMahon functions
\be \label{RefinedM}
M_\delta (x , q , \sqrt{ \kappa}) = \prod_{i,j}^{\infty} \left( 1 -  x \, q^{i+j-1} \sqrt{ \kappa}^{i-j+\delta} \right)^{-1} \, ,
\ee
where for future use we have generalize the MacMahon function to depend on an extra parameter $x$. With this notation, $x=1$ and $\delta=+1$ is the choice of \cite{NO} while the choice of $x = 1$ and $\delta=0$ is the one made in \cite{DG}. 

A concrete way to obtain the refined limit is for example to choose a subtorus
\be
t_1 \longrightarrow  c_1 \, x^{N-1} \qquad t_3 \longrightarrow  c_3 \, x \qquad t_2 \longrightarrow c_2 \, 1/x^N
\ee
with $N \in \zed$ very large, and then send $x \longrightarrow 0$. The parameters ${c_1,c_2,c_3}$ are chosen to enforce the condition $t_1 t_2 t_3 = \kappa$. With this choice we get
\be
Z^{ref}_{\complex^3} (q ; \kappa) = 1 - q \ \sqrt{ \kappa} + q^2  (1 + 2  \kappa ) - q^3 \frac{1 + 2  \kappa + 3  \kappa^2}{\sqrt{ \kappa}}  + \cdots
\ee
which agrees with the first terms of the refined MacMahon function
$
M_{\delta=1}  (1, q,  \sqrt{ \kappa} )
$.
Finally the cohomological limit is obtained by sending $\kappa \longrightarrow 1$.

The $\complex^3$ partition function can be in principle generalized to any toric manifold, constructed by gluing together copies of $\complex^3$. To each copy one can associate a vertex with boundary conditions given by the asymptotics at infinity of a plane partition. The full vertex is not known in closed form except in particular cases \cite{Kononov:2019fni}. The partition function can however be obtained order by order. 

In this perspective the results of \cite{NO} correspond to the lift to M-theory of the BPS state counting problem formulated in the large radius chamber. In the rest of the article we will extend this construction to a different chamber.

\section{Noncommutative algebraic geometry}

In this Section we quickly review some basic elements of noncommutative algebraic geometry in the context of local Calabi-Yau threefolds. Certain toric singularities admit crepant resolutions which can be understood as moduli spaces of representations of a noncommutative algebra, the Jacobi algebra of a quiver with superpotential \cite{vandenbergh}. Such an algebra is an example of a noncommutative crepant resolution. We will briefly explain the relation between such resolutions and quivers and then discuss their role in defining noncommutative Donaldson-Thomas theory. The reader is referred to the reviews \cite{Cirafici:2018jor,Cirafici:2012qc} for a more in depth discussion.

\subsection{Quivers and noncommutative crepant resolutions}

Several aspects of BPS invariants on local threefolds can be understood algebraically via quiver representation theory \cite{szendroi,Cirafici:2010bd,Cirafici:2011cd,Cirafici:2008sn}. A quiver is a finite directed graph identified by the two finite sets $\mathsf{Q}_0$ and $\mathsf{Q}_1$, the nodes and the arrows, and by the two maps $s,t \, : \, \mathsf{Q}_1 \longrightarrow \mathsf{Q}_0$ that to an arrow $a \in \mathsf{Q}_1$ associate its starting vertex $s (a) \in \mathsf{Q}_0$ and its terminal vertex $t (a) \in \mathsf{Q}_0$.

The algebra of paths $\mathbb{C} \mathsf{Q}$ is the unital associative $\mathbb{C}$--algebra spanned by all paths in the quiver with product given by concatenation of paths if possible and zero otherwise. The Jacobian algebra $\mathsf{A} = \mathbb{C} \mathsf{Q} / \langle \mathsf{R} \rangle$ is the algebra of paths modulo the ideal generated by a set of relations $\mathsf{R}$, given as $\complex$--linear combinations of paths in $\mathbb{C} \mathsf{Q}$. We will be interested in the case where the relations $\mathsf{R}$ are derived from a superpotential.  A superpotential is an element of the vector space of cyclic monomials $\mathsf{W} \in \mathbb{C} Q / \left[  \mathbb{C} Q, \mathbb{C} Q \right]$,
and the ideal of relations is given by $\mathsf{R} = \langle \partial_a \mathsf{W} \, \vert \, a \in \mathsf{Q}_1 \rangle$.

Representations of a quiver are defined by assigning a complex vector space $V_i$ to each node $i \in \mathsf{Q}_0$ and a morphism $B_a \in \mathrm{Hom} (V_{s(a)} , V_{t(a)})$ to each arrow $a \in \mathsf{Q}_1$. The dimension vector of the representation has components $d_i = \mathrm{dim} V_i $. The representation space is defined by fixing a dimension vector $d \in \mathbb{N}^{\sfQ_0}$ as
\be \label{RepSpace}
\mathrm{Rep}_d (\sfQ) = \bigoplus_{a \, : \, i \longrightarrow j} \Hom_\complex (V_i , V_j) \, .
\ee
We denote by $\mathsf{rep} (\mathsf{Q})$ the category of quiver representations. If the quiver has a superpotential we require the morphisms to be compatible with the relations $\partial \, \mathsf{W} = 0$. Now $\mathrm{Rep}_d (\sfQ , \sfW) $ denotes the subscheme of \eqref{RepSpace} obtained by imposing the relations $\partial \, \sfW = 0$. Similarly one can introduce the category of representations of a quiver with superpotential $\mathsf{rep} (\mathsf{Q} , \mathsf{W})$. This category is equivalent to the category $\mathsf{A}-\mathsf{mod}$ of left $\mathsf{A}$-modules. In most physical applications the objects of interest are isomorphism classes of representations; that is the orbits with respect to the action of the group $G_d = \prod_{i \in \mathsf{Q}_0} \, GL (V_i , \mathbb{C})$. In this case we are led to the quotient stack $\mathsf{M}_d (\sfQ) = \mathrm{Rep}_d (\sfQ , \sfW) / G_d$.

In many physical situations one is interested in framed quivers and their representations. A framed quiver can be obtained for example by adding an extra vertex $\{ \bullet \}$ together with an additional arrow $a_{\bullet}$ such that $s (a_{\bullet}) = \bullet$ and $t (a_{\bullet}) = v_0$, where $v_0 \in \mathsf{Q}_0$ is a reference node of $\mathsf{Q}$. The new quiver $\widehat{\mathsf{Q}}$ has $\widehat{\mathsf{Q}}_0 = \mathsf{Q}_0 \cup \{ \bullet \}$ and $\widehat{\mathsf{Q}}_1 = \mathsf{Q}_1 \cup \{ \bullet \xrightarrow{a_\bullet} v_0 \}$. Similarly one can define the Jacobian algebra of a framed quiver, framed representations and their moduli spaces.

Simple modules $\mathsf{D}_v$ are associated to each vertex $v \in \sfQ_0$ and have $V_v = \complex$ and $V_w = 0$ for $w \neq v$. Physically such modules describe fractional branes. Let $\sfe_v$ be
the trivial path at $v$ of length zero. Then $\sfP_v: =
\sfA \, \sfe_v$ is the subspace of the path algebra which is generated
by all paths that begin at vertex $v$. These are projective objects in
the category ${\sf{rep}}(\sf Q , W)$ . Then a projective resolution of the simple module $\mathsf{D}_v$ is given by
\begin{equation}
\begin{xy}
\xymatrix@C=8mm{  \cdots \ \ar[r] & \ \displaystyle{\bigoplus_{w\in
      {\sfQ_0}} \, \sfP_w{}^{\oplus d^{k}_{w,v}}} \ \ar[r] & \ \cdots \ \ar[r]
  & \ \displaystyle{\bigoplus_{w \in {\sfQ_0}} \, \sfP_w{}^{\oplus d^{1}_{w,v}}
  }  \ \ar[r] & \ar[r]  \ \sfP_v \ & \ \sfD_v \ \ar[r]  & \ 0
}
\end{xy}
\label{projressimple}\end{equation}
where
\begin{equation}
{d}^p_{w,v} = \dim \Ext^p_{\sf{A}} \left( \sfD_v , \sfD_w \right) \ .
\label{dpwvdimExt}\end{equation}
In particular $d^0_{w,v}=\delta_{w,v}$ since $\sfD_v$ are simple objects;
 $d^1_{w,v}$ gives the number of arrows in $\sfQ_1$ from
vertex $w$ to vertex $v$; $d^2_{w,v}$ gives the number of relations expressed as sum of paths from
vertex $w$ to vertex $v$;
$d^3_{w,v}$ is the number of relations among the relations, and
so on.

The path algebra of a quiver can be thought of as a noncommutative analog of a polynomial algebra, while quiver representations are a noncommutative analog of coherent sheaves. In the case of local threefolds this can be made more precise. The relevant object which describes BPS states on the threefold is $D^b (\mathrm{coh} (X))$, the homotopy category of complexes of sheaves where we invert all quasi-isomorphisms. Complexes of sheaves represent branes wrapping cycles in the Calabi-Yau geometry, and different complexes which give rise to the same physical state have to be identified.

If $X$ arises as a crepant resolution of a singularity, then it also admits a noncommutative crepant resolution \cite{vandenbergh}. Crepant resolutions and noncommutative crepant resolutions are derived equivalent. The noncommutative resolution is a certain algebra $\sfA$ for which $D^b (\mathrm{coh} (X)) = D^b (\sfA - \rm{mod})$. 

In this paper we will limit ourselves to toric Calabi-Yau $X$, for which $D^b (\mathrm{coh} (X))$ is equivalent to the bounded derived category of representations of a quiver with superpotential. This is precisely the quiver which captures the low energy description of supersymmetric bound states of D-branes probing the singularity. In this case the equivalence is induced by a tilting sheaf $\mathscr{T} = \bigoplus \mathscr{T}_i$, whose summands represent elementary D-branes. A tilting sheaf is a generator of $D^b (\mathrm{coh} (X))$ (in the sense that the smallest triangulated full subcategory of $D^b (\mathrm{coh} (X))$ containing all the summands $\{ \mathscr{T}_i \}$ is $D^b (\mathrm{coh} (X))$), with the property that  $\mathrm{Ext}^i (\mathscr{T} , \mathscr{T}) = 0$ for $i>0$, and such that $\mathsf{A} = \mathrm{End} (\mathscr{T})$ has finite global dimension.

In particular the algebra $\mathsf{A}$ can be written as the path algebra of a certain quiver with relations: this quiver has nodes corresponding to the summands $\mathscr{T}_i$ and the arrows corresponding to morphisms between the $\mathscr{T}_i$. Its center $\sfZ  (\sfA)$ is the coordinate ring of the singularity and its moduli spaces of representations are resolutions of the singularity. For a recent review about these statements as well as references to the original literature, we refer the reader to \cite{Mozgovoy:2020has}.

%
%

\subsection{Noncommutative Donaldson-Thomas theory}

On a noncommutative crepant resolution we can define BPS states. The low energy effective theory of a D2--D0 system is described by a supersymmetric quiver quantum mechanics with superpotential. Geometrically the quiver is
derived through the endomorphisms of a tilting object. To incorporate configurations with non-vanishing D6-brane charge we consider framed quivers. 

The representation theory of quivers as summarized above can be easily extended to framed quivers. More generically one can introduce a framing node with many arrows going out and incoming from the unframed quiver. If  such arrows form new loops one can introduce a framed superpotential $\hat{\sfW}$. In this note we will not consider this case (for us it will be $\sfW = \hat{\sfW}$), but it will be relevant to the study of wall-crossing. Finally, we will also only consider framed representations where the framing node is one dimensional. 

The representations of this new quiver can be constructed as before, by specifying a $d_v$-dimensional vector space $V_v$ on each node $v\in\sfQ_0$, while the framing node $\bullet$ always carries a one-dimensional vector space $\IC$. This determines a representation space, and its subscheme $\mathrm{Rep} (\hat{\sfQ} ,  \sfW , v_0)$ cut out by the relations $\partial \, \sfW = 0$. By taking the quotient with respect to the group $G_d$ (which does not involve the framing node), we obtain $\mathsf{M}_d (\hat{\sfQ} , v_0)$, the moduli space of framed representations with framing at the node $v_0$, with dimension vector $d=(d_v)_{v\in\sfQ_0}$. One can also introduce a framed dimension vector $\hat{d} = (1,d)$.

Physically D6-D2-D0 BPS states are identified with stable representations of $\hat{\sfQ}$. One interprets the simple representations as elementary BPS constituents which interact via the superpotential $\hat{\sfW}$. When a representation of $(\hat{\sfQ} , \sfW)$ with dimensions $d_i = \dim \, V_i$ exists and is stable, this is interpreted physically as the existence of a bound state of charge $\gamma = \gamma_{D6} +  \sum_i \, d_i \, \gamma_i$. Here $\gamma_{D6}$ represents the D6 brane charge, while all the other $\gamma_i$ are associated to the nodes of the unframed quiver. The stability condition is determined by a central charge function, a linear map $Z  \, : \, K (\mathsf{rep} (\hat{\sfQ} , \sfW)) \longrightarrow \complex$, with the property that it takes values in the upper half plane for every elementary charge associated to the nodes of the quiver. We say that a state of charge $\gamma_r$ described by a representation $\mathsf{R} \in \mathsf{rep} (\hat{\sfQ} , \sfW)$ is stable if for any proper sub-representation $\mathsf{S} \in \mathsf{rep} (\hat{\sfQ} , \sfW)$ associated to a state of charge $\gamma_s$, we have that $\arg Z (\mathsf{S} ) < \arg Z (\mathsf{R} )$ (at a point in the moduli space). For simplicity we will assume that we can chose the moduli in such a way that the inequality is strict, so that we don't have to consider semi-stable representations.

In the case at hand, one can find a region of the moduli space where the only stable representations are cyclic. This is discussed in depth in \cite{Chuang:2013wt} and here we will only highlight the main points. The cyclicity condition can be equivalently stated by saying that a representation $\mathsf{R}$ is cyclic if any non-zero sub-representation with non-trivial support at the framing node, that is dimension vector $\hat{d} = (1,d)$, is $\mathsf{R}$ itself. This region can be identified by taking carefully the limit where the mass of the non-compect D6 brane goes to infinity. Schematically, the central charge of a bound state $\mathsf{R}$ of dimension vector $\hat{d} = (1,d)$ can be written as
\be
Z (\mathsf{R}) = \xi + \sum_i d_i z_i
\ee 
by linearity. The stability condition can be rephrased in terms of the slope of $\mathsf{R}$
\be
\mu (\mathsf{R}) = - \frac{\mathrm{Re (Z (\mathsf{R}) )} }{\mathrm{Im} (Z (\mathsf{R}) )} \, .
\ee
in the sense that the condition $\arg Z (\mathsf{S} ) < \arg Z (\mathsf{R} )$ above is equivalent to $\mu (\mathsf{S} ) < \mu (\mathsf{R} )$, see for example \cite{Mozgovoy:2020has} for a discussion. If we write the central charge of the D6 brane as $\xi = |\xi| \e^{\ii \phi}$ and then send $|\xi| \longrightarrow \infty$ while keeping all the other $|z_i|$ fixed, then we see that for any framed representation $\mathsf{R}$
\be
\mu (\mathsf{R}) = - \frac{|\xi| \cos \varphi + \mathrm{Re} ( \sum_i d_i z_i) }{|\xi| \sin \varphi + \mathrm{Im} ( \sum_i d_i z_i)} \longrightarrow - \cot \varphi
\ee  
Since $\mathsf{R}$ had dimension vector $\hat{d} = (1,d)$, any sub-representation $\mathsf{S}$ can have dimension vector $\hat{d} = (0,d)$ or $\hat{d} = (1,d)$. This means that the stability condition for the framed representation $\mathsf{R}$ can be rephrased by requiring  $\mu (\mathsf{S} ) < - \cot \varphi$ for any sub-representation with dimension vector $\hat{d} = (0,d)$, and $\mu (\mathsf{S}' ) > - \cot \varphi$ for any non-trivial quotient $\mathsf{R} \twoheadrightarrow \mathsf{S}'$ with dimension vector $\hat{d} = (0,d)$. The reason quotient representations appear is to deal more easily with sub-representations with $\hat{d} = (1,d)$ in the $|\xi| \longrightarrow \infty$ limit above.

Finally one can go to an asymptotic chamber where $ - \cot \varphi >> 0$. In such a chamber sub-representations with $\hat{d} = (0,d)$ are not destabilizing; on the other hand quotient representations with  $\hat{d} = (0,d)$ are ruled out. This implies that the only sub-representation of $\mathsf{R}$ with non-trivial support on the framing node is $\mathsf{R}$ itself. This is precisely the cyclic stability condition. 

In this chamber, sometimes called the noncommutative resolution chamber, the stable representations correspond to cyclic modules of the framed quiver. We will denote by $\scrM_{d}(\hat\sfQ,v_0)$, or sometimes just by $\scrM_d$ the moduli space of cyclic modules. Such a moduli space is well-behaved ~\cite{reineke} and enjoys a symmetric perfect obstruction theory~\cite{szendroi,reineke} which follows from the general results of \cite{fantechi}. It is in general non-compact. In such a chamber one can defined the noncommutative Donaldson-Thomas invariants as the {weighted} Euler characteristic of these moduli spaces as
\begin{equation} \label{ncdt}
\DT_{d} (\sfA) := \chi \big(\scrM_d \,,\, \nu_\sfA
\big) \ ,
\end{equation}
which physically represents the index of BPS bound states constituted by a single D6 brane and a gas of D2-D0 states. Here $\nu_\sfA$ is the Behrend function of $\sfA$ \cite{kai}, a constructible function with the property that the weighted Euler characteristic \eqref{ncdt} gives the degree of the virtual fundamental class of $\sfA$.
%
For these invariants we define a cohomological partition function associated with the algebra $\sfA$
\begin{equation}
Z^{coho}_{\sfA} (p ) = \sum_{d_v \in \zed} \,
\DT_{d}(\sfA) \ \prod_{v\in\sfQ_0}\, p_v^{d_v} \ .
\label{NCZBPS}\end{equation}
Here, we have introduced a $|\sfQ_0|$ dimensional vector $p$ to keep track of the nodes upon which the representations are based. This enumerative problem can be studied via a combinatorial algorithm obtained by applying the virtual localization formula on the representation moduli space, with respect to a certain natural toric action. The action of the torus $\torus$ is defined by rescaling each arrow in a way compatible with the $\partial \, W = 0$ equations, modulo a sub-torus contained in $G_d$; we will be more precise in the following Sections. The fixed points of this action are ideals in the path algebra $\sfA$. They admit a graphical representation in terms of certain combinatorial arrangements known as \textit{pyramid partitions}.

It follows from a result by Behrend-Fantechi \cite{fantechi} that for a symmetric perfect obstruction theory of the kind constructed in
 \cite{szendroi,reineke} the contribution of an isolated fixed point to the
Euler characteristic (\ref{ncdt})
is just a sign determined by the parity of the dimension of the tangent space
at the fixed point. Therefore if all fixed points $\pi$ are isolated we obtain
\beq
\DT_{d} (\sfA) = \sum_{\pi\in\scrM_{d}^\torus} \, (-1)^{\dim T_{\pi}
  \scrM_{d}} \ .
\label{DTfantechi}\eeq
While this formula has a concrete combinatorial description, at first sight its physical interpretation is less clear because one loses track of the relation between the D-brane charges defined at large radius and the combinatorial objects. The physical states at the singularity have charges which are non-trivial functions of the D0 and D2 brane charges at large radius. The relevant dynamical variables at the singularities should be interpreted as fractional branes, stuck at the singularity. This change of variables is known in some models \cite{Cirafici:2010bd}. 

\section{The noncommutative conifold}

In this Section we will study the K-theoretic DT theory for the conifold quiver. We will briefly review the relevant quiver and its representation theory and then compute the M2-brane index by a direct localization computation. We then provide a conjecture for the full partition function and discuss some of its limits.

\subsection{Geometry and BPS states}

We will now consider the conifold geometry. At large radius it is given by the total space of the bundle $X = \cO (-1) \oplus \cO (-1) \longrightarrow \PP^1$. In this phase the topological string captures the BPS invariants corresponding to bound states of a single D6 brane with a gas of D0/D2 branes, with a strong B-field \cite{Iqbal:2003ds}. In this limit the problem of computing BPS invariants corresponds to the study of Donaldson-Thomas invariants associated with the Hilbert scheme of points and curves $\mathrm{Hilb} (X ; n , \beta)$, with $n$ the number of $D0$ branes and $\beta$ the class of the curve wrapped by the $D2$ branes. 

We will be interested in the singular limit where the size of the $\PP^1$ is sent to zero. In the noncommutative phase the conifold geometry is replaced by the representation theory of the Klebanov-Witten quiver
\be
\xymatrix{ 
& \circ \ar@/^/[rr]^{a_1 , a_2} & & \bullet \ar@/^/[ll]^{b_1 , b_2}
}
\ee
with superpotential $\sfW =  a_1 b_1 a_2 b_2 - a_1 b_2 a_2 b_1$. We will denote by $\complex \sfQ$ the algebra of paths, defined as the algebra generated by the arrow set, with product given by the concatenation of arrows whenever possible. The Jacobian path algebra is $\sfA = \complex \sfQ / \langle \partial \sfW \rangle$. The center of the path algebra $\sfZ (\sfA)$ is spanned by the variables
\begin{align}
x_1 &= a_1 b_1 + b_1 a_1 \cr
x_2 &= a_2 b_2 + b_2 a_2 \cr
x_3 &= a_1 b_2 + b_2 a_1 \cr
x_4 &= a_2 b_1 + b_1 a_2 
\end{align}
so that ${\rm Spec} \, R = {\rm Spec} \, \complex[x_1,x_2,x_3,x_4] / (x_1 x_2 - x_3 x_4)$ is the ring of functions on the conifold singularity. Therefore the path algebra of the Klebanov-Witten quiver sees the conifold singularity as its center. The path algebra itself is a noncommutative crepant resolution. In this case the equivalence $\mathsf{D}^b (\mathsf{A-mod}) \simeq \mathsf{D}^b (\mathrm{coh} \, X)$ is determined by the tilting bundle $\cO \oplus \pi^* \cO_{\PP^1} (1)$, where $\cO$ is the structure sheaf on the resolved conifold and $\pi$ is the natural projection on the $\PP^1$; see \cite{nagao,Nagao:2010kx} for a discussion in this context. 

We would like to understand BPS states in the noncommutative phase. The study of BPS states for the conifold quiver was done in \cite{szendroi}. To model D-brane bound states one adds a framing node to the quiver, corresponding to a single D6 brane
\be
\xymatrix{ \star \ar[dr] & & & \\
& \circ \ar@/^/[rr]^{a_1 , a_2} & & \bullet \ar@/^/[ll]^{b_1 , b_2}
}
\ee
BPS invariants associated with bound states of D-branes now correspond to noncommutative Donaldson-Thomas invariants associated with this quiver \cite{szendroi}. We want to generalize this computation to the K-theory setting.

\subsection{Toric action and fixed points}

Consider the representation space
\be
\mathrm{Rep} (\sfQ , v_\circ) = \bigoplus_{(v \rightarrow w) \in \sfQ_1} \Hom_\complex (V_v , V_w) \oplus \Hom (\complex , V_\circ ) \, .
\ee
The last factor is associated to the framing and determines a vector $v_\circ \in V_\circ$. A suitable stability condition which selects the noncommutative crepant resolution chamber is given by only considering \textit{cyclic modules}, that is modules which are generated by $v_\circ$ \cite{szendroi,Cirafici:2012qc}. Now we consider the open subvariety $\mathrm{Rep}^c (\sfQ , v_\circ)$ defined by the condition that the module is cyclic. Similarly we can consider the representation space $\mathrm{Rep} (\sfQ , v_\circ ; \mathsf{W})$ defined by imposing the F-term relations $\partial \, \sfW = 0$ on $\mathrm{Rep} (\sfQ , v_\circ)$, and the corresponding subvariety of cyclic modules $\mathrm{Rep}^c (\sfQ , v_\circ ; \mathsf{W})$. Finally the relevant moduli space is
\be
\mathscr{M}_{d_\circ,d_\bullet} = \left[ \mathrm{Rep}^c (\sfQ , v_\circ ; \mathsf{W}) \Bigg/ GL(d_\circ , \complex) \times GL (d_\bullet , \complex) \right] \, .
\ee
We will denote a (representative of an isomorphism class of a) cyclic module as $(v_\circ , M)$.

This moduli space carries a natural toric action which can be used to classify fixed points. The classification is carried out in Proposition 2.5.1 of \cite{szendroi}; since the proof is constructive, we summarize it here in some detail. The only difference is that we do not impose that the determinant of the toric action is trivial. We have the following tori
\begin{itemize}
\item $\torus_F$. This is the natural torus which rescales each arrow by a phase, which we will denote by $\mathrm{diag} \left( t_{a_1} , t_{a_2} , t_{b_1} , t_{b_2} \right)$. Note that this torus also preserves the F-term relations and therefore acts naturally on the space of cyclic modules $\mathrm{Rep}^c (\sfQ , v_\circ ; \mathsf{W})$.
\item $\torus_G$. This is the ``gauge'' torus, induced by the $GL (d_\circ , \complex) \times GL (d_\bullet , \complex)$ action which changes the bases of $V_\circ$ and $V_\bullet$. This group contains a $(\complex^*)^2$ factor, whose elements $(\mu_\circ , \mu_\bullet) \in (\complex^*)^2$ act on representation maps by conjugation, say $\mu_\circ X_a \mu_\bullet^{-1}$ or  $\mu_\bullet X_b \mu_\circ^{-1}$. However the induced gauge torus is just $\torus_G = \complex^* = (\complex^*)^2 / \complex^*$, since diagonal elements act trivially. We take this torus acting by weights $(-1,-1,1,1)$ (since it must act on the $b$'s in the opposite way as on the $a$'s). This torus however does not influence the fixed point classification.
\item $\torus_{F , \sfW}$. This is the sub-torus of $\torus_F$ with the weight condition $t_{a_1} t_{a_2} t_{b_1} t_{b_2} = \kappa$. In \cite{szendroi} the condition is $\kappa = 1$
\item Finally the torus $\torus_\sfW = \torus_{F , \sfW} / \torus_G$. 
\end{itemize}

Now we proceed to study the fixed points of $\mathscr{M}_{d_\circ,d_\bullet} $. The result of \cite{szendroi} is that such fixed points correspond to ideals in $\sfA$. To begin with, a cyclic module $(v_\circ , M)$ is canonically associated to an ideal (its annihilator) $I$, as $M = \sfA / I$. In particular since $v_\circ$ is based at the vertex $\circ$, this ideal has the form $I = I_\circ \oplus P_\bullet$, where $P_\bullet = \sfA \e_\bullet$ are the paths starting from the vertex $\bullet$.

Note that $\torus_\sfW = \torus_{F , \sfW} / \torus_G$-fixed points of $\mathscr{M}_{d_\circ,d_\bullet} $ correspond to $\torus_{F , \sfW}$-fixed ideals $I$ in $\sfA$. The reason is that the generators of $I_\circ$ are sum of path algebra monomials with the same starting point and endpoint. Therefore on each generator the gauge torus acts by multiplication by a constant. Since $I_\circ$ is an ideal, it does not change under this operation. 

\begin{figure}[H]
\centering
\def\svgwidth{8cm}
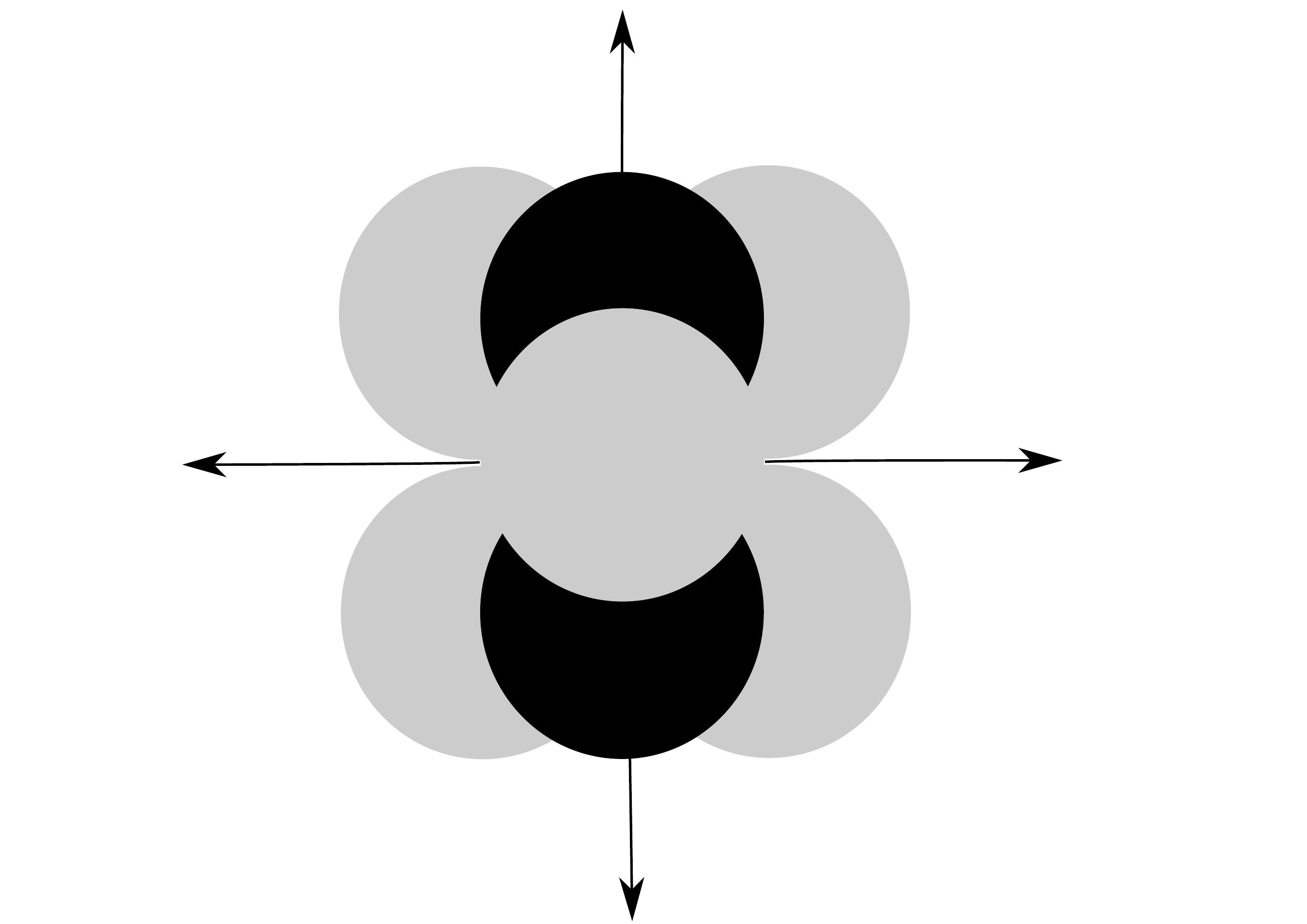
\caption{Pyramid arrangement. We show the tip of an infinite pyramid; white and black stones are associated with the two nodes of the Klebanov-Witten quiver. One moves from one stone to another via the arrows of the quiver; the toric weight of the arrows is shown explicitly. }
\label{PyramidArrangement}
\end{figure}

\begin{figure}[H]
\centering
\def\svgwidth{8cm}
\includegraphics{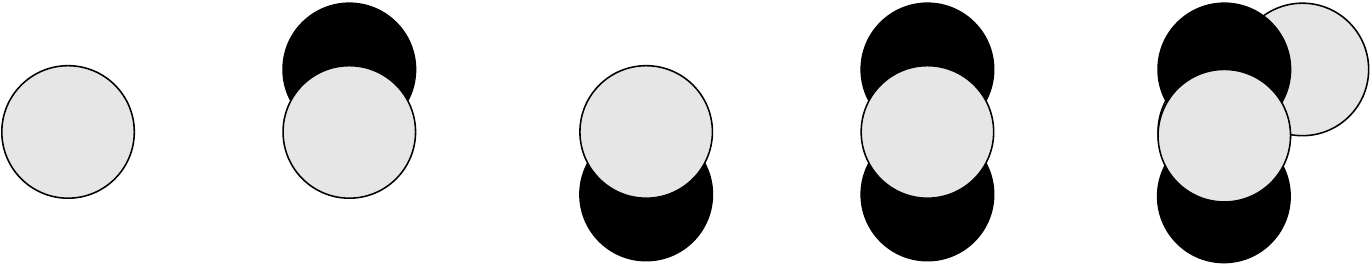}
\caption{Conifold fixed points. Each fixed point corresponds to a pyramid partition, a configuration of stones which can be consistently removed from the pyramid arrangement.}
\label{conifoldFP}
\end{figure}

It is shown in \cite{szendroi} that such ideals are in one to one correspondence with pyramid partitions. Pyramid partitions are defined as follows. First we define a pyramid arrangement, which is a three dimensional infinite pyramid; the tip of this pyramid is shown in Figure \ref{PyramidArrangement}. The first layer (starting from the top of the pyramid) of this configuration corresponds to the vector $v_\circ$, the second layer corresponds to the two ways we can reach the node $\bullet$ starting from $\circ$, that is $a_1 v_\circ$ and $a_2 v_\circ$, and so on. The shape of the pyramid partition is determined by the superpotential relations $\partial \, \sfW = 0$.
Given a pyramid arrangement, one defines a pyramid partition $\pi$ by the following condition: $\pi$ is a configuration of stones such that for each stone in $\pi$, the ones immediately above it (of a different color) are in $\pi$ as well. Note that this is basically the same definition of a Young diagram, which guarantees that the complement of a pyramid partition is an ideal. Examples of pyramid partitions are shown in Figure \ref{conifoldFP}.
Now we argue that pyramid partitions correspond to fixed points. 

First consider a module $M$. By cyclicity, it is spanned by vectors of the form 
\be \label{spanningset}
\{ v_\circ , a_1 v_\circ , a_2 v_\circ , b_1 a_1 v_\circ , b_2 a_1 v_\circ , \dots \}.
\ee
From this set of vectors we remove those which are linearly dependent when imposing the F-term relations $\partial \, \sfW = 0$. Then to each of the remaining vectors, which now form a basis, we associate a stone in a pyramid partition. The partition has the correct form because of the F-term equations.

Conversely, consider a partition $\pi$. To each stone we assign a basis vector $v_{i,a}$, with $a= \circ , \bullet$. Then we collect all the $v_{i,a}$ with the same color for $a$ and use them as a spanning set to construct a vector space $V_a$, whose dimension is the number of stones with that color. The arrows of the quiver induce maps between these vectors, which obey the relations $\partial \, \sfW = 0$ by construction. Finally the corresponding module is $M = \complex \oplus V_\circ \oplus V_\bullet$, where the first factor corresponds to the framing node.

This completes the classification of fixed points in terms of pyramid partitions \cite{szendroi}. Note that to compute the K-theoretic partition function it is not enough to enumerate the combinatorial configurations, since we need explicitly the isotypical decomposition of the vector spaces $V_\circ$ and $V_\bullet$ into toric weight spaces. This information can be obtained, for example, from the spanning set \eqref{spanningset}. This is why we have summarized the discussion of \cite{szendroi} in such detail.

\subsection{Localization}

To each fixed point $\pi$ we associate the decomposition in weight spaces
\begin{align}
V_\circ &= \sum_{(i,j,k,l) \in \pi_\circ} \ t_{a_1}^{-i} \, t_{a_2}^{-j} \, t_{b_1}^{-k} \, t_{b_2}^{-l} \\
V_\bullet &= \sum_{(i,j,k,l) \in \pi_\bullet} \ t_{a_1}^{-i} \, t_{a_2}^{-j} \, t_{b_1}^{-k} \, t_{b_2}^{-l}
\end{align}
where the two sets $(\pi_\circ , \pi_\bullet)$ run over stones of the respective color. The weights are those of the generators of the cyclic module, as explained above.

To use the localization formula we need the deformation/obstruction theory around each fixed point. This is given by the virtual tangent space. The virtual tangent space can be constructed in terms of the tangent bundle to $\mathscr{M}_{d_\circ, d_\bullet}$. The tangent bundle has the form ``arrows'' - ``gauge transformations'':
\be
T _\pi = (t_{a_1} + t_{a_2}) \otimes \overline{V}_\circ \otimes V_\bullet + (t_{b_1} + t_{b_2}) \otimes \overline{V}_\bullet \otimes {V}_\circ + V_{\circ} - \overline{V}_\circ \otimes V_\circ - \overline{V}_\bullet \otimes V_\bullet
\ee
where the ``arrows'' carry their toric weight, and we do not consider gauge transformations on the framing node. As in \eqref{Vbar} we have introduced the involution $\overline{t}_i = t_i^{-1}$, so that $\overline{V}_\circ = \sum_{(i,j,k,l) \in \pi_\circ} \ t_{a_1}^{i} \, t_{a_2}^{j} \, t_{b_1}^{k} \, t_{b_2}^{l} $ and $\overline{V}_\bullet = \sum_{(i,j,k,l) \in \pi_\bullet} \ t_{a_1}^{i} \, t_{a_2}^{j} \, t_{b_1}^{k} \, t_{b_2}^{l}$. Then we have\footnote{
While we have checked this directly in the following examples, we have not proven in full generality that the virtual tangent space is movable, that is that it does not contains terms which are fixed by the toric action. It would be important to do so, as it would put our conjectures on a firmer ground.}
\begin{align}
T^{\rm vir}_\pi &= \left( T -  \kappa \otimes T^* \right)_{\pi}
\\
&= (t_{a_1} + t_{a_2}) \otimes \overline{V}_\circ \otimes V_\bullet + (t_{b_1} + t_{b_2}) \otimes \overline{V}_\bullet \otimes {V}_\circ + V_{\circ} - \overline{V}_\circ \otimes V_\circ - \overline{V}_\bullet \otimes V_\bullet
\cr & \nonumber \
- t_{a_1} t_{a_2} t_{b_1} t_{b_2} \left( 
(t_{a_1}^{-1} + t_{a_2}^{-1}) \otimes V_\circ \otimes \overline{V}_\bullet + (t_{b_1}^{-1} + t_{b_2}^{-1}) \otimes V_\bullet \otimes \overline{V}_\circ + \overline{V}_{\circ} - \overline{V}_\circ \otimes V_\circ - \overline{V}_\bullet \otimes V_\bullet
\right)
\end{align}
with $\kappa = t_{a_1} t_{a_2} t_{b_1} t_{b_2}$. 

We define our partition function as 
\be
 \chi (\mathscr{M} , \widehat{\cO}^{\rm vir}) = \sum_{(d_\circ , d_\bullet) } \, q_0^{d_\circ} q_1^{d_\bullet} \, \chi \left( \mathscr{M}_{d_\circ,d_\bullet} ,  (\sign) \cO^{\rm vir}_{\mathscr{M}_{d_\circ,d_\bullet}} \otimes \cK_{\rm vir}^{1/2} \right)
\ee
Here $(\sign)$ refers to the fact that in defining the prefactor of the symmetrized virtual structure sheaf we have the freedom of choosing a sign. We will choose $\sign = (-1)^{d_\circ} $;  but this is just a convention. For the moment we will drop this sign and reinstate it only at the end. We have defined $\cK_{\rm vir}^{1/2} = \det^{-1/2} T^{\rm vir}$ and the full moduli space is given by the disjoint union $\mathscr{M} = \cup_{d_\circ , d_\bullet} \mathscr{M}_{d_\circ, d_\bullet}$.

After expanding and cancelling terms, the virtual tangent space assumes schematically the form $T^{\rm vir} = {\rm Def} - {\rm Obs} = \sum_i w_i - \sum_i \kappa / w_i$, where $w_i$ denotes a generic toric weight. Then the localized virtual structure sheaf has the form
\begin{align}
\cO^{\rm vir}_{\mathscr{M}_{d_\circ , d_\bullet}} \otimes \cK_{\rm vir}^{1/2}  &= \sum_{|\pi_\circ| = d_\circ , |\pi_\bullet| = d_\bullet}  \hat{\sfa} (T^{vir}_{\pi_\circ , \pi_\bullet}) \, \cO_{I_{\pi_\circ, \pi_\bullet}} \cr
& =  \sum_{|\pi_\circ| = d_\circ , |\pi_\bullet| = d_\bullet} \prod_i  \ahat{w_i} \, \cO_{I_{\pi_\circ, \pi_\bullet}}
\end{align}
in K-theory. The sum runs over all the fixed points with $d_\circ$ white stones and $d_\bullet$ black stones. The role of the twist by $\cK_{\rm vir}^{1/2}$ is to transform the usual K-theory weights into $\hat{\sfa} (T^{\rm vir})$.

\subsection{Some explicit fixed points}

Here we collect some explicit results of the localization computation. Higher order computations become quickly rather technical. The interested reader can find some of them in the supporting \textsc{mathematica} files \cite{mathe}. 

\begin{enumerate}
\item Fixed point with $d = (d_\circ , d_\bullet) = (1,0)$. This is the first fixed point in Figure \ref{conifoldFP}. It has $V_\circ = 1$ and $V_\bullet = \emptyset$. Then $T^{\rm vir} = 0$, which means that the deformations exactly cancel the obstructions. We have $\hat{\sfa} (T^{\rm vir}) = 1 $.

\item Fixed points with $d= (1,1)$. There are two fixed points, corresponding to the representations with $V_\circ=1$, $V_\bullet =1/ t_{a_1}$ and $V_\circ = 1$, $V_\bullet =1/ t_{a_2}$, as $\torus_\sfW$-modules. The corresponding pyramid partitions are shown in Figure \ref{conifoldFP}. 
In the first case 
\be
T^{\rm vir}_1 = \frac{t_{a_2}}{t_{a_1}}+t_{a_1} t_{b_1}-t_{a_2} t_{b_1}+t_{a_1}
   t_{b_2}-t_{a_2} t_{b_2}-t_{a_1}^2 t_{b_1} t_{b_2}
\ee
which contributes
\be
\begin{gathered}
\hat{\sfa} (T^{\rm vir}_1) = \ahat{t_{a_2} / t_{a_1}} \, \ahat{ t_{a_1} t_{b_1}} \, \ahat{ t_{a_1} t_{b_2}}  \, .
\end{gathered}
\ee
The second fixed point is related to the first by the exchange $a_1 \leftrightarrow a_2$.
%
%
Summing the two contributions gives
\be
\left( -\frac{(\kappa t_{a_1}-t_{a_2}) (t_{a_1} t_{b_1}-\kappa) (t_{a_1}
   t_{b_2}-\kappa)}{\kappa^{3/2} (t_{a_1}-t_{a_2}) (t_{a_1} t_{b_1}-1) (t_{a_1}
   t_{b_2}-1)}-\frac{(t_{a_1}-\kappa t_{a_2}) (t_{a_2} t_{b_1}-\kappa) (t_{a_2}
   t_{b_2}-\kappa)}{\kappa^{3/2} (t_{a_1}-t_{a_2}) (t_{a_2} t_{b_1}-1) (t_{a_2}
   t_{b_2}-1)} \right) 
\ee

\item Fixed point with $d=(1,2)$. There is only one, shown in Figure \ref{conifoldFP}, with $V_\circ = 1$ and $V_\bullet =1/ t_{a_1} +1/ t_{a_2}$. We see that $T^{\rm vir} = 0$, so its contribution is just $\hat{a} (T^{\rm vir}) = 1$ 

\item Fixed points with $d=(2,1)$. There are four fixed points, which we write as tuples $\{ V_\circ , V_\bullet \}$ 
\begin{align}
\left\{\frac{1}{t_{a_1} t_{b_1}}+1,\frac{1}{t_{a_1}}\right\} \qquad 
\left\{\frac{1}{t_{a_1} t_{b_2}}+1,\frac{1}{t_{a_1}}\right\} \qquad
\left\{\frac{1}{t_{a_2} t_{b_1}}+1,\frac{1}{t_{a_2}}\right\} \qquad
\left\{\frac{1}{t_{a_2} t_{b_2}}+1,\frac{1}{t_{a_2}}\right\}
\end{align}
The first one corresponds to the virtual tangent space
\be
T^{\rm vir}_1 =
 \frac{t_{a_2}}{t_{a_1}}+t_{a_1} t_{b_1}-t_{a_1} t_{a_2}
   t_{b_1}^2-t_{a_2} t_{b_2}+\frac{t_{b_2}}{t_{b_1}}-t_{a_1}^2
   t_{b_1} t_{b_2} \, ,
\ee
while the other can be obtained from this one by permuting the toric weights.
%
%
The full contribution of the four fixed points is
\begin{align}
-\frac{(\kappa t_{a_1}-t_{a_2}) (-\kappa+t_{a_1} t_{b_1}) (\kappa
   t_{b_1}-t_{b_2})}{\kappa^{3/2} (t_{a_1}-t_{a_2}) (-1+t_{a_1} t_{b_1})
   (t_{b_1}-t_{b_2})}
   -\frac{(t_{a_1}-\kappa t_{a_2}) (-\kappa+t_{a_2} t_{b_1})
   (\kappa t_{b_1}-t_{b_2})}{\kappa^{3/2} (t_{a_1}-t_{a_2}) (-1+t_{a_2} t_{b_1})
   (t_{b_1}-t_{b_2})}
   \cr
   -\frac{(\kappa t_{a_1}-t_{a_2}) (-t_{b_1}+\kappa t_{b_2})
   (-\kappa+t_{a_1} t_{b_2})}{\kappa^{3/2} (t_{a_1}-t_{a_2}) (-t_{b_1}+t_{b_2})
   (-1+t_{a_1} t_{b_2})}
   -\frac{(t_{a_1}-\kappa t_{a_2}) (-t_{b_1}+\kappa
   t_{b_2}) (-\kappa+t_{a_2} t_{b_2})}{\kappa^{3/2} (t_{a_1}-t_{a_2})
   (-t_{b_1}+t_{b_2}) (-1+t_{a_2} t_{b_2})} \, .
\end{align}

\item Fixed points with $d=(2,2)$. There are eight fixed points (one of which is shown in Figure \ref{conifoldFP}), we list them in the form $\{ V_\circ , V_\bullet \}$
\begin{small}
\begin{align}
& \left\{\frac{1}{t_{a_1} t_{b_1}}+1,\frac{1}{t_{a_1}^2
   t_{b_1}}+\frac{1}{t_{a_1}}\right\}
   \, 
\left\{\frac{1}{t_{a_1} t_{b_2}}+1,\frac{1}{t_{a_1}^2
   t_{b_2}}+\frac{1}{t_{a_1}}\right\}
   \,
   \left\{\frac{1}{t_{a_2} t_{b_1}}+1,\frac{1}{t_{a_2}^2
   t_{b_1}}+\frac{1}{t_{a_2}}\right\}
   \cr
&  \left\{\frac{1}{t_{a_2} t_{b_2}}+1,\frac{1}{t_{a_2}^2
   t_{b_2}}+\frac{1}{t_{a_2}}\right\}
   \,
   \left\{\frac{1}{t_{a_1}
   t_{b_1}}+1,\frac{1}{t_{a_1}}+\frac{1}{t_{a_2}}\right\}
   \,
   \left\{\frac{1}{t_{a_1}
   t_{b_2}}+1,\frac{1}{t_{a_1}}+\frac{1}{t_{a_2}}\right\}
   \cr
& \left\{\frac{1}{t_{a_2}
   t_{b_1}}+1,\frac{1}{t_{a_1}}+\frac{1}{t_{a_2}}\right\}
   \,
   \left\{\frac{1}{t_{a_2}
   t_{b_2}}+1,\frac{1}{t_{a_1}}+\frac{1}{t_{a_2}}\right\}
\end{align}
\end{small}

The overall contribution is somewhat involved and can be found in the supporting \textsc{mathematica} file \cite{mathe}.
\item Fixed points with $d = (3,1)$. They are
\begin{align}
V_\circ & = 1 + 1/ (t_{a_1} t_{b_1}) + 1/ (t_{a_1} t_{b_2}) \qquad V_\bullet = 1/ t_{a_1} \cr
V_\circ & = 1 + 1/ (t_{a_2} t_{b_1}) + 1/ (t_{a_2} t_{b_2}) \qquad V_\bullet = 1/ t_{a_2} 
\end{align}

\item Fixed points with $d = (2,3)$. They are
\begin{align}
V_\circ & = 1 + 1/ (t_{a_1} t_{b_2}) \qquad V_\bullet = 1/ t_{a_1} + 1/(t_{a_1}^2 t_{b_2}) + 1/ t_{a_2} \cr
V_\circ & = 1 + 1/ (t_{a_1} t_{b_1}) \qquad V_\bullet = 1/ t_{a_1} + 1/(t_{a_1}^2 t_{b_1}) + 1/ t_{a_2} \cr
V_\circ & = 1 + 1/ (t_{a_2} t_{b_2}) \qquad V_\bullet = 1/ t_{a_2} + 1/(t_{a_2}^2 t_{b_2}) + 1/ t_{a_1} \cr
V_\circ & = 1 + 1/ (t_{a_2} t_{b_1}) \qquad V_\bullet = 1/ t_{a_2} + 1/(t_{a_2}^2 t_{b_1}) + 1/ t_{a_1} 
\end{align}

\end{enumerate}
and so on. The rest of the fixed points up to $(d_\circ , d_\bullet ) = (4,4)$ can be found in the supporting \textsc{mathematica} file \cite{mathe}.

\subsection{A conjecture for the full partition function}

We can use the partial knowledge given by the few fixed points listed above to propose a conjecture for the full K-theoretic partition function. As explained above we will keep open the freedom to chose the sign of the counting variables.  We define the following function associated to the noncommutative conifold
\begin{align} \label{FNCC}
 \cF_{NCC} [t_{a_1} , t_{a_2} , t_{b_1} , t_{b_2} , t_4 , t_5 , q_1] = &
\frac{\frac{1}{q_1}+q_1}{\left(-\frac{1}{\sqrt{t_4}}+\sqrt{t_4}\right)
   \left(-\frac{1}{\sqrt{t_5}}+\sqrt{t_5}\right)}
   \\ &+ \cF_{\complex^3} [ t_{a_1} t_{b_1}, t_{a_1} t_{b_2} , \frac{t_{a_2}}{t_{a_1}} ,t_4,t_5]  + \cF_{\complex^3} [ t_{a_2} t_{b_1} , t_{a_2} t_{b_2} , \frac{t_{a_1}}{t_{a_2}},t_4,t_5] \, . \nonumber
\end{align}
In this formula it is understood that $\kappa = t_{a_1} t_{a_2} t_{b_1} t_{b_2}$. The variable $q_1$ plays the role of a K\"ahler modulus. The factor $q_1 + \frac{1}{q_1}$ captures the ``K\"ahler dependent'' part of the partition function (meaning that this is the analog of the K\"ahler modulus in the noncommutative phase).

The remaining terms are precisely copies of the $\complex^3$ partition function $\cF_{\complex^3} [t_1,t_2,t_3,t_4,t_5]$, with the substitutions
\begin{align}
t_1 \longrightarrow t_{a_1} t_{b_1} \, , \qquad t_2 \longrightarrow t_{a_1} t_{b_2} \, , \quad t_3 \longrightarrow \frac{t_{a_2}}{t_{a_1}} \, , \cr
t_1 \longrightarrow t_{a_2} t_{b_1} \, , \qquad t_2 \longrightarrow t_{a_2} t_{b_2} \, , \quad t_3 \longrightarrow \frac{t_{a_1}}{t_{a_2}} \, .
\end{align}
These two copies differ by $t_{a_1} \leftrightarrow t_{a_2}$. Note that these two sets of weights match those acting on the resolved conifold $\cO (-1)^{\oplus2} \longrightarrow \PP^1$. The resolved conifold can be covered with two toric charts, and if we let the first set act in one toric chart, the second set then acts on the other.  In particular since these two charts are copies of $\complex^3$, the contribution
\be
\mathsf{S}^\bullet \left( 
\cF_{\complex^3} [ t_{a_1} t_{b_1}, t_{a_1} t_{b_2} , \frac{t_{a_2}}{t_{a_1}} ,t_4,t_5]  + \cF_{\complex^3} [ t_{a_2} t_{b_1} , t_{a_2} t_{b_2} , \frac{t_{a_1}}{t_{a_2}},t_4,t_5]
\right)
\ee
corresponds to the K-theoretic Donaldson-Thomas invariants of points on the resolved conifold.

The conjecture is inspired by the cohomological answer \cite{szendroi}
\be
Z_{NCC}^{coho} (q, z) = M (-q)^2 \prod_{k \ge 1} \left( 1- (-q)^k z \right)^k \left( 1 - (-q)^k z^{-1} \right)^k
\ee
with $q = q_0 q_1$ and $z = q_1$. The ``K\"ahler" dependent part has the $q_1 + 1/q_1$ structure. The remaining part is two copies of the MacMahon function.

We define the full partition function by symmetrization
\begin{align}
\sfZ_{NCC} \left( t_{a_1} , t_{a_2} , t_{b_1} , t_{b_2} , t_4 , t_5 , q_1 \right) &= \sfS^\bullet \cF_{NCC} [t_{a_1} , t_{a_2} , t_{b_1} , t_{b_2} , t_4 , t_5 , q_1] \cr &= \exp \sum_{r \ge 1} \frac{1}{r} \cF_{NCC} [t_{a_1}^r , t_{a_2}^r , t_{b_1}^r , t_{b_2}^r , t_4^r , t_5^r , q_1^r] 
\end{align}

Using \textsc{mathematica} \cite{mathe}, we arrive at the following
\begin{conjecture} \label{conjConifold}
Upon specialization of the parameters $t_4$ and $t_5$ the symmetrized partition function is the generating function of K-theoretic Donaldson-Thomas invariants
\begin{align}
\sfZ_{NNC} \left(\{ t_i \} , q_0 q_1 \kappa^{-1/2} , (q_0 q_1)^{-1} \kappa^{-1/2} , q_1 \right) &=  \sum_{(d_\circ , d_\bullet) } \, (-q_0)^{d_\circ} q_1^{d_\bullet} \, \chi \left( \mathscr{M}_{d_\circ,d_\bullet} ,  \cO^{\rm vir}_{\mathscr{M}_{d_\circ,d_\bullet}} \otimes \cK_{\rm vir}^{1/2} \right) \, .
\end{align}
\end{conjecture}
In the above conjecture we have reintroduced the sign in the prefactor of the virtual structure sheaf; different conventions will just change the signs of the counting parameters. The substitution $t_4 = q_0 q_1 \kappa^{-1/2}$ and $t_5 =  (q_0 q_1)^{-1} \kappa^{-1/2}$ is the same as in $\complex^3$ with counting parameter $q = q_0 q_1$. The compact form of \eqref{FNCC} allows us easily to switch between sign conventions.

\subsection{Refined and unrefined limits}

To take the refined limit we scale the toric weights to zero or infinity, in such a way that the product $\kappa$ remains constant \cite{NO}. When taking this scaling limit 
the contribution of each fixed point becomes
\be
\hat{\sfa} (\sum_i w_i -  \kappa \sum_i w_i^{-1} ) \longrightarrow (-  \kappa^{1/2})^{Index} 
\ee
where 
\be
Index = \# \{i \vert w_i \longrightarrow 0 \} - \# \{ i \vert w_i \longrightarrow \infty \}
\ee
is the number of weights which go to zero minus the number of weights which go to infinity. In our case we take the torus where $t_{a_1}$, $t_{a_2}$ and $t_{b_1}$ go to zero, the first very fast and the latter two very slowly, while $t_{b_2}$ goes to infinity fast, such that the product of the toric weights is a constant\footnote{To be more concrete, we take the subtorus $\{ t_{a_1} \rightarrow c_{1} \, x^{1000} , t_{a_2} \rightarrow c_{2} \, x^{10} , t_{b_1} \rightarrow c_{3} \, x^{10} , t_{b_2} \rightarrow c_{4} \, 1/ x^{1020} \}$, where the parameters $c_i$ have been chosen to ensure that $t_{a_1} t_{a_2} t_{b_1} t_{b_2} = \kappa$, and then send $x \longrightarrow 0$. The specific form of the result will depend on the chosen subtorus.}. 

To compute the refined limit we can use Conjecture \eqref{conjConifold} by taking the refined limit directly. Define the function
\begin{align} \label{FNCCref}
F_{NCC}^r [q_0 , q_1 , \kappa] 
=-\frac{q_0 \left(-\sqrt{\kappa} \left(q_1^2+1\right)+\kappa
   q_1+q_1\right)}{\left(\sqrt{\kappa}-q_0 q_1\right) \left(\sqrt{\kappa}
   q_0 q_1-1\right)} \, . \end{align}
This function arises from taking the refined limit of \eqref{FNCC}. Then conjecture \eqref{conjConifold} implies that 
\begin{align} \label{NCCref}
& Z_{NCC}^{ref} (q_0 , q_1 ; \kappa) :=  \sum_{(d_\circ , d_\bullet) } \, q_0^{d_\circ} q_1^{d_\bullet} \, \chi^{ref} \left( \mathscr{M}_{d_\circ,d_\bullet} ,  \cO^{\rm vir}_{\mathscr{M}_{d_\circ,d_\bullet}} \otimes \cK_{\rm vir}^{1/2} \right) = \sfS^\bullet F_{NCC}^r [-q_0 , q_1 , \kappa] \, .
\end{align}
Note that this result is explicitly invariant for $\kappa \leftrightarrow 1/\kappa$ due to the form of \eqref{FNCCref}. With this sign conventions the refined degeneracies have the form of characters for the parameter $\sqrt{\kappa}$.

To compare with the literature, we need to introduce a bit of notation, to take into account the multiple refinements of the MacMahon function, as well as different signs in the conventions. Define \cite{MMNS}
\begin{align}
G_{re} [a_0 , a_1] &= \prod_{j=0}^{a_0 -1} \left( 1 - \kappa^{- \frac{a_0}{2} + \frac12 + j} (- q_0)^{a_0} q_1^{a_1} \right) 
\end{align}
and consider the function
\be
G (\epsilon , \delta) = \prod_{i \ge 1} G_{re} [i , i-1]  G_{re} [i-1 , i]  M_{2 \epsilon -1} (1 , -q_0 q_1 , \sqrt{\kappa}) M_{2 \delta -1} (1 , -q_0 q_1 , \sqrt{\kappa}) 
\ee
where the refined MacMahon function was introduced in \eqref{RefinedM}. Using this notation the result \eqref{NCCref} is $G (\epsilon = 0 , \delta = 1)$;  we have checked this identification with \textsc{mathematica} and we conjecture it to hold to all orders.
This should be compared with other results in the literature. For example the refined noncommutative conifold partition function of \cite{DG} can be written as $G (1/2 , 1/2)$, while the motivic partition function computed in \cite{MMNS} is $G (1,2)$. Interestingly our partition function agrees with a theorem of \cite{MMNS} according to which the product of the refined MacMahon factors in front of the partition function does not have the form of a square of a single refined MacMahon function. It also agrees with the physical expectation that such partition functions should have coefficients which are linear combinations of $SU(N)$ characters. Note however that the differences between the partition functions can all be traced to the ambiguity of defining the MacMahon function, arising from the non-compactness of the moduli spaces, see \cite{Mozgovoy:2020has} for a recent discussion. Furthermore our result depends explicitly on the torus chosen to take the refined limit. Different tori will give a priori different results. We won't pursue the issue here further, and hope to return to a more systematic study of the refined limits in the future.
%

%
%
%
%
In the unrefined limit, the noncommutative partition function has the following relation with the large radius partition functions \cite{szendroi}
\be
Z_{NCC}^{coho, red} (q,z) = Z_{resolved}^{coho, red} (q,z) \, Z_{resolved}^{coho,red} (q, \frac1z)
\ee
where the reduced partition functions are obtained by removing the MacMahon factors. From this we are lead to conjecture an analogous relation between the reduced K-theoretic partition functions. In particular this produces a prediction for the large radius conifold partition function, as 
\be
 Z'_{DT} (t_4 , t_5 , q_1) = \sfS^\bullet \frac{q_1}{\left(-\frac{1}{\sqrt{t_4}}+\sqrt{t_4}\right)
   \left(-\frac{1}{\sqrt{t_5}}+\sqrt{t_5}\right)}
\ee
which agrees with the result of \cite{Kononov:2019fni}, upon the usual change of variables $t_4 = q / \sqrt{\kappa}$ and $t_5 = 1/ (q \sqrt{\kappa})$. A more systematic analysis of the relation between the noncommutative chamber and the other chambers will appear elsewhere \cite{MWC}.

\section{Orbifold Singularities}

Now we will move to orbifold singularities in toric Calabi-Yau threefolds and set up a formalism to compute the K-theoretic Donaldson-Thomas invariants. The counting of BPS states for orbifold singularities is obtained in terms of quivers associated to the singularity via the three dimensional McKay correspondence. From this quiver we set up the localization computation as in the previous sections. For singularities of the form $\complex^3 / \Gamma$ with $\Gamma \subset \mathrm{SL(3,\complex)}$ a finite subgroup which lies in the maximal torus of $\mathrm{SL (3 , \complex)}$, the formalism reduces to an appropriate $\Gamma$ projection of the $\complex^3$ case. The results can be interpreted as membrane contributions to the M-theory index at the singularity. We will obtain closed forms of the partition function for certain $\Gamma$. The results in this section are based on and generalize those of \cite{Cirafici:2010bd,Cirafici:2011cd}

\subsection{BPS states and McKay quivers}

In this Section we will focus on BPS countings associated with quivers which arise from orbifold singularities of the form $\mathbb{C}^3 / \Gamma$ where $\Gamma \subset \mathrm{SL} (3 , \complex)$ is a finite subgroup which lies in the maximal torus of $ \mathrm{SL} (3 , \complex)$. In this case the relevant quiver is the McKay quiver $\sfQ_\Gamma$ which is constructed out of the representation theory data of $\Gamma$ as follows. The set of vertices of $\sfQ_\Gamma$ is in one to one correspondence with the irreducible one-dimensional representations $\rho_a$ of $\Gamma$. Such representations form a group which we will denote by $\hat{\Gamma}$.

The arrow structure of $\sfQ_\Gamma$ is dictated by the tensor product decomposition: the number of arrows going from node $\rho_b$ to node $\rho_a$ is given by 
\begin{equation} \label{tensordecomp}
a^{(1)}_{ba}=\dim_\mathbb{C} \mathrm{Hom}_\Gamma\big(\rho_b , Q \otimes \rho_a \big) \, ,
\end{equation}
where $Q = \rho_{a_1} \oplus \rho_{a_2} \oplus \rho_{a_3}$ denotes the fundamental three-dimensional representation of $\Gamma$. This is defined by the action of $\Gamma$ on $\complex^3$. The condition that $\Gamma \subset \mathrm{SL(3,\mathbb{C})}$ implies that $a_1+a_2+a_3 = 0$. This construction is the higher dimensional generalization of the ordinary McKay quiver for ALE singularities.

Consider now a representation of the McKay quiver. From the individual vector spaces $V_a$ assigned to the vertices labelled $\rho_a$ we form $V = \bigoplus_{a\in \hat{\Gamma}}\, V_a \otimes \rho^\vee_a$ (here $\rho^\vee$ is the conjugate representation). In other words the representations of the McKay quiver have a natural $\Gamma$-module structure. The linear maps between vector spaces are determined by $B \in \mathrm{Hom}_{\Gamma} (V , Q \otimes V)$. If we introduce the notation $B_\alpha^{(a)} \in \mathrm{Hom}_\mathbb{C} (V_{a}, V_{a+ a_\alpha})$, the relations between the quiver maps can be derived from the superpotential
\begin{equation}
\mathsf{W}_{\Gamma}= \sum_{a\in\hat{\Gamma}}\, 
B_1^{(a+a_2+a_3)}\,\Big( B_2^{(a+a_3)}\, B_3^{(a)}-B_3^{(a+a_2)}\,
  B_2^{(a)} \Big) \ .
\end{equation}
The Jacobian algebra $\mathsf{A}_\Gamma = \mathbb{C} \mathsf{Q}_\Gamma / \langle \partial \, \mathsf{W}_\Gamma \rangle$ is a noncommutative crepant resolution of the singularity $\mathbb{C} / \Gamma$. 

As showed in  \cite{Cirafici:2010bd}, from the point of view of the six dimensional topological theory living on the D6 brane the dimensions of the vector spaces $\mathrm{dim} V_a = k_a$ which enter in the isotypical decomposition $V = \bigoplus_{a\in \hat{\Gamma}}\, V_a \otimes \rho^\vee_a$, correspond to instanton configurations which transform in the irreducible representation $\rho_a$. Similarly in the framing node decomposition $W=\bigoplus_{a\in\widehat\Gamma}\, W_a\otimes \rho_a^\vee$ the vector spaces label boundary conditions at infinity. Since the gauge connection is flat at infinity, it is labelled by an irreducible representation $\rho$ of $\Gamma$. In this note we will only consider trivial representations at infinity: the framing node is connected to the quiver only to the vertex associated with the trivial representation.

The formation of D0/D2 bound states can be seen from the point of view of the quiver quantum mechanics which arises from the quantization of the collective coordinates around each instanton configuration; or more precisely from a direct parametrization of the moduli space of ideal sheaves  \cite{Cirafici:2010bd}. This quiver quantum mechanics localizes onto fixed points of its BRST operator. This operator can be twisted by the natural toric action on $\complex^3$, since the $\torus^3$ action and the $\Gamma$ action commute. Fixed points are again classified by plane partitions $\pi$, except that now they carry a $\Gamma$ action. The reason for this is that the superpotential $\mathsf{W}_\Gamma$ can be seen as the same as the superpotential $\mathsf{W}$ where one takes into account also the $\Gamma$-action on the representation spaces $\Hom_\complex (V , V)$. As a consequence the classification of the fixed points in terms of 3d partitions remains the same, except that now each box also carries a $\Gamma$-weight, corresponding to the (inverse of the) weight of the monomial in $\complex [z_1 , z_2 , z_3 ]$ represented by that box. 

While working equivariantly with respect to the $\Gamma$ action on $\complex^3$, one also inherits the stability condition of the Donaldson-Thomas problem on $\complex^3$. This selects as stable modules  the cyclic modules generated starting from the framing node, in our case the one corresponding to the trivial representation $\rho_0$. On $\complex^3$ this stability condition enters in the classification of the fixed points given in terms of 3d partitions: these are constructed starting from the framing node in a fashion consistent with the equations of motion $\left[B_i , B_j \right]= 0$ for $i,j=1,2,3$. Therefore the classification of fixed points in the $\complex^3 / \Gamma$ case automatically follows from the stability condition. 

\subsection{Localization on quiver moduli spaces}

The relevant quiver representation moduli space associated with the low energy effective quantum mechanics of the D6-D0 system is the quotient stack
\begin{equation}
\mathscr{M}_{\Gamma} (\mathbf{v},\mathbf{w}) = [ \mathrm{Rep}^c (\mathsf{\hat{Q}}_{\Gamma} ; \mathsf{W}_{\Gamma}) / \prod_{r\in\widehat\Gamma}\, GL(k_r,\mathbb{C}) ] \, ,
\end{equation}
where $\mathrm{Rep}^c (\mathsf{\hat{Q}}_{\Gamma} ; \mathsf{W})$ is the subset of
\begin{equation}
\mathrm{Hom}_{\Gamma} (V , Q \otimes V) \ 
 \oplus \ \mathrm{Hom}_{\Gamma} (V , \mbox{$\bigwedge^3$} Q \otimes V) \ 
\oplus \ \mathrm{Hom}_{\Gamma} (W , V) \ 
\end{equation}
cut out by the F-term equations derived from the superpotential $\mathsf{W}_{\Gamma}$ and parametrizing cyclic representations\footnote{To be more precise the middle term $ \mathrm{Hom}_{\Gamma} (V , \mbox{$\bigwedge^3$} Q \otimes V) $ corresponds to an extra field $\varphi$ which arises in the effective quantum mechanics of the D6-D0 bound states on $\complex^3$. This field does not enter in the geometric description of $\mathrm{Hilb}^n (\complex^3)$. If one denotes by $I$ the field corresponding to $ \mathrm{Hom}_{\Gamma} (\complex , V) $, then the equation $I^\dagger \varphi = 0$ is required to make \eqref{equivdefcomplex} an equivariant complex. We refer the reader to \cite{Cirafici:2008sn,Cirafici:2010bd} for a more complete discussion.}. Here $\mathbf{v} = \{ v_a \}_{a \in \hat{\Gamma}}$ where $v_a = \dim_\complex V_a$. As explained above, we can assume $W = \complex$ and therefore $\mathbf{w} = (1, 0, \dots , 0)$. Therefore we will remove it from the notation and denote the moduli spaces by $\mathscr{M}_\Gamma (\mathbf{v})$.

%

In instanton counting problems one can construct a local model for the moduli space by studying a deformation complex, which parametrizes linearized field configurations near solutions of the equations of motion, up to gauge invariance. In the case of orbifold singularities it is given at a fixed point by \cite{Cirafici:2010bd}
\begin{equation} \label{equivdefcomplex}
\xymatrix{
  \mathrm{Hom}_{\Gamma}  (V_{\pi} , V_{\pi})
   \quad\ar[r] &\quad
   {\begin{matrix}  \mathrm{Hom}_{\Gamma} (V_{\pi} , V_{\pi} \otimes Q )
   \\ \oplus \\
    \mathrm{Hom}_{\Gamma} (\complex , V_{\pi}) \\ \oplus  \\  \mathrm{Hom}_{\Gamma} (V_{\pi} ,
   V_{\pi}  \otimes \bigwedge^3 
   Q) \end{matrix}}\quad \ar[r] & \quad
   {\begin{matrix}  \mathrm{Hom}_{\Gamma} (V_{\pi} , V_{\pi}  \otimes \bigwedge^2
       Q) \\ \oplus \\ 
        \mathrm{Hom}_{\Gamma} (V_{\pi},\complex \otimes \bigwedge^3 Q)
   \end{matrix}}
} \, .
\end{equation}
The complex \eqref{equivdefcomplex} is a way to encode the information about the virtual tangent space which is alternative to the construction discussed in Section \ref{M2braneindex}. If we forget about the $\Gamma$ action, the above complex becomes precisely the relevant complex for $\complex^3$, whose space of configurations is given by $\mathrm{Hilb} (\complex^3, n)$. Also note that from the perspective of the effective D6-D0 quantum mechanics, the term $ \mathrm{Hom}_{\Gamma} (V_{\pi} , V_{\pi}  \otimes \bigwedge^3 Q)$ appears as a field in the complex \eqref{equivdefcomplex} , while in \eqref{TvirC3}, is part of the obstruction bundle, but with a minus sign, signalling an overcounting of obstructions.

From the deformation complex we can read directly the virtual tangent space
\begin{equation} \label{orbcharacter}
( T^{\rm vir}_\pi)^\Gamma = \big( V_{\pi} -
\overline{V_{\pi}} \ t_1 t_2 t_3 - (1-t_1)\, (1-t_2)\,
(1-t_3) ~ \overline{V}_{\pi} \otimes V_{\pi} \big)^{\Gamma} \ ,
\end{equation}
which becomes the virtual tangent space \eqref{TvirC3} for the case of $X = \complex^3$ when we forget about the $\Gamma$-action. In \eqref{orbcharacter}, and in all the remaining examples in this Section, we have used the involution $\overline{t} = t^{-1}$ to define $ \overline{V}_{\pi} $. Fixed points of the toric action are still classified by 3d partitions, since the orbifold group $\Gamma$ is a subgroup of the torus $\torus^3$. The vector space $V$ can be decomposed at a fixed point $\pi$ as
\begin{equation}
V_{\pi} =\bigoplus_{a\in\widehat{\Gamma}} \,  \left( P_{l,a} \otimes \rho_a^{\vee} \right) 
\label{VpiGammadecomp}\end{equation}
The modules $P_{l,a}$ are associated to the decomposition of $\sum_{(n_1,n_2,n_3)\in \pi_l}\, t_1^{-n_1+1} \,t_2^{-n_2+1}\,t_3^{-n_3+1} $ as a  $\Gamma$-module, while the $\rho_a^\vee$ factors keep track of the representation. The vector spaces $V$ have a decomposition as a sum of monomials in the toric variables, and each monomial has a definite transformation under the orbifold action. Each 3d partition carries an action of the group $\Gamma$ induced by the fundamental representation $Q = \rho_{a_1} \oplus \rho_{a_2} \oplus \rho_{a_3}$. In plain words each box of the partition is associated to a character of $\Gamma$ depending on the transformation of the monomial in the toric variables associated with that box under the $\Gamma$ action.

We define the K-theoretic BPS partition function as
\be
\sfZ_\Gamma ( \{ p_i \} , \{ t_i \}) = \chi (\mathscr{M}_\Gamma , \widehat{\mathscr{O}}_\Gamma^{\rm vir})
\ee
where additional counting variables $\{ p_i \}$ can be introduced to keep track of the dimensions of the representations by a simple modification of the virtual structure sheaf. The moduli space decomposes as a disjoint union
\be
\mathscr{M}_\Gamma = \bigsqcup \mathscr{M}_\Gamma ( \mathbf{v} ) \, .
\ee
The modified virtual structure sheaf is 
\be
\widehat{\mathscr{O}}_\Gamma^{\rm vir} = \mathrm{prefactor} \ \mathcal{O}_\Gamma^{\rm vir} \otimes \mathcal{K}_{\mathrm{vir}, \Gamma}^{1/2}
\ee
where the prefactor contains counting parameters to label each component of the moduli space, and is a vector of the form $ \prod_{i \in \sfQ_0} p_i^{v_i}$.

As before, all the information about the modified virtual structure sheaf $\widehat{\mathscr{O}}_\Gamma^{\rm vir} $ is contained in $T^{\rm{vir}}_\Gamma$. The virtual structure sheaf parametrizes obstructions while $\mathcal{K}^{1/2}_{\rm{vir}, \Gamma} = \mathrm{det}^{-1/2} T^{\rm vir}_\Gamma$. 

The virtual tangent space can now be interpreted as the $\Gamma$-invariant part of \eqref{TvirC3}
\be
(T_\pi^{\rm vir})^\Gamma = ( T_\pi \widetilde{M} )^\Gamma - (\kappa \otimes T^*_\pi \widetilde{M})^\Gamma = \sum_i a_i - \sum_i b_i = \sum_i w^{\Gamma}_i - \sum \frac{\kappa}{w^{\Gamma}_i}
\ee
where the weights $w^\Gamma_i$ are in general monomials in the toric weights which are \textit{invariant} under the $\Gamma$-action. Note that since $\Gamma$ has trivial determinant, the combination $\kappa = t_1 \, t_2 \, t_3$ is invariant. 

At a fixed point $\pi$ the virtual structure sheaf can be written in localized K-theory as
\be
( \cO_\pi^{\rm vir} )^{\Gamma} = \cO_\pi \prod_i \frac{1-b_i^{-1}}{1-a_i^{-1}} \, .
\ee
Similarly
\be
\cK_{{\rm vir},\Gamma} = \frac{\det \left(T^* \widetilde{M} \otimes \kappa \right)^\Gamma}{\det  \left(T \widetilde{M} \right)^\Gamma}
\ee
Finally, putting everything together (in localized K-theory)
\be
(\widehat{\cO}_\pi^{\rm vir})^\Gamma =  \cO_\pi \prod_i \frac{1- b_i^{-1}}{1-a_i^{-1}} \otimes (\cK^{\rm vir})^{1/2} = \cO_\pi \prod_i \frac{(\kappa / w^\Gamma_i)^{1/2} - (\kappa / w^\Gamma_i)^{-1/2}}{(w^\Gamma_i)^{1/2} - (w^\Gamma_i)^{-1/2}} 
\ee
where again we stress that the weights are $\Gamma$-invariant since $\kappa = t_1 t_2 t_3$ is invariant due to the Calabi-Yau condition.

The refined limit is taken precisely as in the affine space case, by sending the toric weights $t_i$ to zero or infinity while keeping their product $\kappa$ constant. Sending $\kappa \longrightarrow 1$ reproduces the cohomological limit \cite{Cirafici:2010bd}
\be
Z_\Gamma^{coho} (\{ p_i \}) = \sum_{\pi} \prod_{i \in \sfQ_0} p_i^{v_i}   \int_{ [\mathscr{M}_\Gamma (\mathbf{v}) ]^{\rm vir}} \ 1 \, .
\ee

\subsection{The case of A-fibered singularities}

We begin by studying a series of examples of orbifold singularities which are fibrations of ADE singularities over the complex plane. Such singularities have semi-small crepant resolutions which are fibrations of ALE spaces over the complex plane. We will however work in the noncommutative crepant resolution chamber, using quiver representation data to compute BPS partition functions. In this example we will consider the simplest case where $X = \complex^2 / \zed_n \times \complex$ and the singularity is resolved as a noncommutative crepant resolution. In this example we will also denote by $\Gamma$ the orbifold group acting on the $\complex^2$ factor, hoping that this will not cause confusion.


\paragraph{Geometry and representation theory}

We will consider orbifolds of the form $X = \complex^2 / \Gamma \times \complex$ where $g \in \Gamma = \zed_n$ acts on $\complex [z_1 , z_2 , z_3]$ as
\be
g \cdot (z_1 , z_2 , z_3) = (\zeta \, z_1 , \zeta^{-1} \, z_2 , z_3) \, 
\ee
where $\zeta$ is a $n$-th root of unity. At large radius such singularities have semi-small crepant resolutions given by trivial fibration of a resolved $A_{n-1}$ singularity over $\complex$. The resolved geometry is toric and the exceptional locus consists of $-2$ curves which intersect transversally. The intersection matrix of the exceptional curves is the negative of the Cartan matrix of $A_{n-1}$. Tilting sheaves for these geometries where discussed in Section 1.1 of \cite{nagao} using combinatorial tools from toric geometry, see in particular Theorems 1.10 and 1.15.

At short distances smooth geometry is replaced by representation theory data. In our case this is given by the McKay quiver associated to the singularity. The nodes of the quiver are in one to one correspondence with the irreducible representations $\rho_\alpha$ of $\Gamma \subset \mathrm{SU} (2) \subset \mathrm{SU} (3)$. The arrows are given by the multiplicities of the decomposition of $Q \otimes \rho_r$ into irreducible representations, where $Q$ denotes the fundamental representation. 

For example if we take $\Gamma= \zed_3$, the defining representation is $Q = \rho_1 \oplus \rho_2 \oplus \rho_0$. From \eqref{tensordecomp} we now see 
\begin{equation}
a_{rs}^{(1)} = \left(
  \begin{matrix} 1 & 1 & 1 \\ 1 & 1 & 1 \\ 1 & 1 & 1 \end{matrix}
\right) \ .
\end{equation}
The quiver $\mathsf{Q}$ constructed with these data is
\begin{equation}
\vspace{4pt}
\begin{xy}
\xymatrix@C=20mm{
& \ v_0 \ \bullet \ \ar@/^/[ddl] \ar@/_0.5pc/[ddr] \ar@(ur,ul) & \\
& & \\
 \ v_1 \ \bullet \ \ar@/_/[rr] \ar@/^/[uur] \ar@(lu,ld) & & \ \bullet \
v_2 \ \ar@/_/[uul] \ar@/_/[ll]  \ar@(rd,ru)
}
\end{xy}
\vspace{4pt}
\end{equation}
The quiver $\hat{\mathsf{Q}}$ which is relevant to the problem of computing degeneracies of BPS bound states is obtained by framing $\mathsf{Q}$. The framing node represents a single D6 brane and is connected to $\mathsf{Q}$ by a single arrow to the node labelled by $v_0$. From a gauge theory perspective the choice of this node corresponds to the choice of a superselection sector where the gauge connection at infinity is flat and in the trivial representation of $\Gamma$. 

We will first go through a few examples and then make a general conjecture for the partition function for arbitrary $\Gamma= \zed_n$.

\paragraph{BPS partition function for $\complex^2 / \zed_2 \times \complex$}

We begin with a simple example. In this case the $\Gamma$-action on $\complex [z_1 , z_2 , z_3]$ is 
\be
g \cdot (z_1 , z_2 , z_3) = (\zeta z_1 , \zeta z_2 , z_3)
\ee
with $\zeta^2=1$. 

At a fixed point the representation spaces decompose
\be
V = V_0 + V_1 = \sum_{(i,j,k) \in \pi} t_1^{-i+1} t_2^{-j+1} t_3^{-k+1} \delta_{\left( i-j=0 \!\!\!\!\mod 2\right)} + \sum_{(i,j,k) \in \pi} t_1^{-i+1} t_2^{-j+1} t_3^{-k+1} \delta_{\left( i-j=1  \!\!\!\!\mod 2 \right)}
\ee
we set $|\pi_i|= \dim \, V_r$ for $r=0,1$. These measure the number of boxes in the 3d partition which transform in the conjugacy class labelled by $r$. In terms of this decomposition the virtual tangent space has the form
\begin{align}
(T_\pi^{\mathrm vir})^\Gamma =& V_0 - \overline{V}_0 \, t_1 \, t_2 \, t_3 - (V_0 \otimes \overline{V}_0+ V_1 \otimes \overline{V}_1) (1+ t_1 \, t_2) (1 - t_3) 
\cr & - (V_0 \otimes \overline{V}_1 + V_1 \otimes \overline{V}_0) (-t_1 -t_2) (1- t_3) \, .
\end{align}
We can now compute the partition function
\be
\sfZ_{\complex^2 / \zed_2 \times \complex} (p_0 , p_1 , \{ t_i \}) = \sum_{n} \sum_{|\pi| = n} p_0^{|\pi_0|} \, p_1^{|\pi_1|} \, \hat{\mathsf{a}} \left( (T_\pi^{\mathrm{vir}})^\Gamma \right)
\ee
where the second sum is over all three dimensional partitions with fixed number of boxes. We can see directly the first few terms, up to degree two
\begin{align}
\sfZ_{\complex^2 / \zed_2 \times \complex} (p_0 , p_1 , \{ t_i \})  =& \ 1+\frac{p_0 (\kappa-t_3)}{\sqrt{\kappa} (-1+t_3)}+\frac{p_0 p_1
   \left(-\kappa+t_1^2\right) (\kappa t_1-t_2) (\kappa-t_3)}{\kappa^{3/2}
   \left(-1+t_1^2\right) (t_1-t_2) (-1+t_3)} \\ & + \frac{p_0
   p_1 (t_1-\kappa t_2) \left(-\kappa+t_2^2\right) (\kappa-t_3)}{\kappa^{3/2}
   (t_1-t_2) \left(-1+t_2^2\right) (-1+t_3)} 
    +\frac{p_0^2
   (-\kappa+t_3) \left(-\kappa+t_3^2\right)}{\kappa (-1+t_3)^2 (1+t_3)} + \cdots \nonumber \, .
\end{align} 
We conjecture that this partition function has a closed form presentation as the symmetrized form of a certain orbifold generalization of Nekrasov's function. 
Define the function
\begin{align}
\mathscr{F}_{\complex^2 / \zed_2 \times \complex} \left[ t_1,t_2,t_3,t_4,t_5 ; p_1 \right] = \left( p_1 + \frac{1}{p_1} \right) \frac{({\kappa} / t_3)^{1/2}- (t_3 / {\kappa})^{1/2}}{\prod_3^5 (t^{1/2}_i - t_i^{-1/2})} \cr +
\mathscr{F}_{\complex^3} \left[ t_1/t_2,t_2^2,t_3,t_4,t_5 \right] + \mathscr{F}_{\complex^3} \left[ t_1^2,t_2/t_1,t_3,t_4,t_5 \right]
\end{align}
Then we claim that
\be
\sfZ_{\complex^2 / \zed_2 \times \complex} (p_0 , p_1 , \{ t_i \}) = \sfS^\bullet  \mathscr{F}_{\complex^2 / \zed_2 \times \complex} \left[ t_1,t_2,t_3,t_4,t_5 ; p_1 \right]
\ee
with the identifications $\kappa = t_1 t_2 t_3$, $t_4 = q/ \sqrt{\kappa}$, $t_5 = 1/(q \sqrt{\kappa})$; finally after symmetrization we set $q = - p_0 p_1$. This claim is checked order by order in the supporting \textsc{mathematica} file \cite{mathe} up to terms of degree six.

\paragraph{BPS partition function for $\complex^2 / \zed_3 \times \complex$}

Now we consider the orbifold $\complex^2 / \zed_3 \times \complex$ with action
\be
g \cdot (z_1 , z_2 , z_3) = (\zeta z_1 , \zeta^{-1} z_2 , z_3)
\ee
with $\zeta^3=1$. We decompose the representation space as $V = V_0 + V_1 + V_2$ with
\begin{align}
V_0 & =  \sum_{(i,j,k) \in \pi} t_1^{-i+1} t_2^{-j+1} t_3^{-k+1} \delta_{i-j=0 \!\!\!\! \mod 3} \cr
V_1 & =  \sum_{(i,j,k) \in \pi} t_1^{-i+1} t_2^{-j+1} t_3^{-k+1} \delta_{i-j=1 \!\!\!\! \mod 3} \cr
V_2 & =  \sum_{(i,j,k) \in \pi} t_1^{-i+1} t_2^{-j+1} t_3^{-k+1} \delta_{i-j=2 \!\!\!\! \mod 3}
\end{align}
In particular we set $|\pi_r| = \dim_\complex V_r$, effectively obtained by setting the toric weights to 1.


Then the virtual tangent space decomposes as
\begin{align}
(T^{\rm vir}_{\pi})^{\zed_3} =& V_0 -\overline{V}_0 \, t_1 \, t_2 \, t_3 - (V_0 \,\overline{V}_0 + V_1\, \overline{V}_1 + V_2 \,\overline{V}_2) (1 + t_1 \, t_2) (1 - t_3) \cr
& - (V_0\, \overline{V}_1 + V_1\, \overline{V}_2 + V_2 \,\overline{V}_0) (-t_2 ) (1 - t_3) \cr & - (V_0 \,\overline{V}_2 + V_1 \,\overline{V}_0 + V_2 \,\overline{V}_1) (-t_1 ) (1 - t_3)
 \, .  
\end{align}
We can now compute the partition function
\be
\sfZ_{\complex^2 / \zed_3 \times \complex} (\{ p_i \} , \{ t_i \}) = \sum_{n} \sum_{|\pi| = n} p_0^{|\pi_0|} \, p_1^{|\pi_1|} \, p_2^{|\pi_2|} \, \hat{\mathsf{a}} \left( (T_\pi^{\mathrm{vir}})^{\zed_3} \right) \, .
\ee
We can see directly the first few terms
\begin{align}
\sfZ_{\complex^2 / \zed_3 \times \complex} & (\{ p_i \} , \{ t_i \})=  1+\frac{p_0 (\kappa-t_3)}{\sqrt{\kappa} (-1+t_3)}+\frac{p_0 p_1
   (\kappa-t_3)}{\sqrt{\kappa} (-1+t_3)}+\frac{p_0 p_2
   (\kappa-t_3)}{\sqrt{\kappa} (-1+t_3)} +\frac{p_0^2 (-\kappa+t_3) \left(-\kappa+t_3^2\right)}{\kappa
   (-1+t_3)^2 (1+t_3)}
   \cr & +\frac{p_0 p_1 p_2
   \left(-\kappa+t_1^3\right) \left(\kappa t_1^2-t_2\right)
   (\kappa-t_3)}{\kappa^{3/2} \left(-1+t_1^3\right) \left(t_1^2-t_2\right)
   (-1+t_3)}+\frac{p_0 p_1 p_2 \left(t_1^2-\kappa
   t_2\right) \left(\kappa t_1-t_2^2\right) (\kappa-t_3)}{\kappa^{3/2}
   \left(t_1^2-t_2\right) \left(t_1-t_2^2\right)
   (-1+t_3)} 
   \cr & +\frac{p_0 p_1 p_2 \left(t_1-\kappa
   t_2^2\right) \left(-\kappa+t_2^3\right) (\kappa-t_3)}{\kappa^{3/2}
   \left(t_1-t_2^2\right) \left(-1+t_2^3\right)
   (-1+t_3)}+\frac{p_0^2 p_1 (-\kappa+t_3)^2}{\kappa
   (-1+t_3)^2}+\frac{p_0^2 p_2 (-\kappa+t_3)^2}{\kappa
   (-1+t_3)^2} 
   \cr & -\frac{p_0^3 (-\kappa+t_3)
   \left(-\kappa+t_3^2\right) \left(-\kappa+t_3^3\right)}{\kappa^{3/2} (-1+t_3)^3
   (1+t_3) \left(1+t_3+t_3^2\right)} + \cdots \, .
\end{align}

Now we want to recast the partition function as a symmetrized form of an orbifold function.
We define the orbifold function as
\begin{align}
\mathscr{F}_{\complex^2 / \zed_3 \times \complex} \left[ t_1,t_2,t_3,t_4,t_5 ; p_1, p_2 \right] = \left( p_1 + \frac{1}{p_1} + {p_2} + \frac{1}{p_2} + p_1 p_2 + \frac{1}{p_1 p_2} \right) \frac{({ \kappa} / t_3)^{1/2}- (t_3 / {\kappa})^{1/2}}{\prod_3^5 (t^{1/2}_i - t_i^{-1/2})} \cr +
\mathscr{F}_{\complex^3} \left[ t_1^3,t_2/t_1^2,t_3,t_4,t_5 \right] + \mathscr{F}_{\complex^3} \left[ t_1^2/t_2,t_2^2/t_1,t_3,t_4,t_5 \right]+ \mathscr{F}_{\complex^3} \left[ t_1/t_2^2,t_2^3,t_3,t_4,t_5 \right]
\end{align}
with the usual identifications $\kappa = t_1 t_2 t_3$, $t_4 = q/ \sqrt{ \kappa}$, $t_5 = 1/(q \sqrt{ \kappa})$. 
Then we can check order by order that 
\be
\sfZ_{\complex^2 / \zed_3 \times \complex} (\{ p_i \} , \{ t_i \}) = \sfS^\bullet  \mathscr{F}_{\complex^2 / \zed_3 \times \complex} \left[ t_1,t_2,t_3,t_4,t_5 ; p_1 , p_2 \right]
\ee
where after symmetrization we set $q = - p_0 p_1 p_2$.
Note that
\begin{align} \label{invC2Z3}
& \frac13 \sum_{r=0}^2 \frac{1}{(1- w^r t_1) (1-w^{2 r} t_2) (1-t_3)} \cr & = \frac{1}{(1- t_1^3) (1-\frac{t_2}{t_1^2}) (1-t_3)} +
\frac{1}{(1- \frac{t_1^2}{t_2}) (1-\frac{t_2^2}{t_1}) (1-t_3)} +
\frac{1}{(1- \frac{t_1}{t_2^2}) (1-t_2^3) (1-t_3)}
\end{align}
is the generating function of $\zed_3$ invariant monomials in the variables $t_i$. The left hand side gives invariants by construction since it averages the generating function of symmetric monomials in the toric variables, over all the images of the orbifold action. The right hand side follows from a direct computation. It can also be interpreted as $\chi (\widetilde{\complex^2 / \zed_3} \times \complex , \cO) $ computed via localization. In this case \eqref{invC2Z3} can be interpreted as the equality $\chi (\complex^2 / \zed_3 \times \complex , \cO) = \chi (\widetilde{\complex^2 / \zed_3} \times \complex , \cO) $. This is the origin of the three copies of the $\complex^3$ generating function above.

\paragraph{General formula}

These results lead us to conjecture a general formula for arbitrary $\complex^2 / \zed_n \times \complex$. Consider first the decomposition
\begin{align}
\frac1n \sum_{r=0}^{n-1} \frac{1}{(1- w^r t_1) (1-w^{- r} t_2) (1-t_3)} = \frac{1}{1-t_3} \sum_{k=0}^{n-1} \frac{1}{(1-\frac{t_1^{n-k}}{t_2^{k}})(1-\frac{t_2^{k+1}}{t_1^{n-1-k}})}
\end{align}
where $w = \exp \frac{2 \pi \ii}{n}$ is an $n^{th}$ root of unity. We define the function
\be
\mathscr{F}_{\complex^2 / \zed_n \times \complex} (t_1,t_2,t_3,t_4,t_5) = \sum_{k=0}^{n-1} \mathscr{F}_{\complex^3} \left[ \frac{t_1^{n-k}}{t_2^{k}} ,  \frac{t_2^{k+1}}{t_1^{n-1-k}} , t_3 , t_4 , t_5 \right]
\ee
which generates $\zed_n$-invariant monomials in the toric parameters. Note that by taking the symmetrization $\mathsf{S}^\bullet$ one finds the K-theoretic Donaldson-Thomas invariants of points of the resolution $\widetilde{\complex^2 / \zed_n} \times \complex$.

Introduce now the variables $p_{[r,s]} = p_r p_{r+1} \cdots p_s$. We define the reduced function
\begin{align}
\mathscr{F}_{\complex^2 / \zed_n \times \complex}^{(n)} \left[ \{ p_i \} , t_1 , t_2 , t_3 , t_4 , t_5 \right] = \sum_{0 < r \le s < n} \left( p_{[r,s]} + \frac{1}{p_{[r,s]}} \right)  \frac{({ \kappa} / t_3)^{1/2}- (t_3 / { \kappa})^{1/2}}{\prod_{i=3}^5 (t^{1/2}_i - t_i^{-1/2})}
\end{align}

We claim the following
\begin{conjecture} 
\begin{align}
\sfZ_{\complex^2 / \zed_n \times \complex} \left( \{ p_i \} , \{ t_i \} \right) = \sfS^\bullet \left( \mathscr{F}_{\complex^2 / \zed_n \times \complex} \left[ t_1,t_2,t_3,t_4,t_5 \right] + \mathscr{F}_{\complex^2 / \zed_n \times \complex}^{(n)} \left[ \{ p_i \} , t_1 , t_2 , t_3 , t_4 , t_5 \right]
\right)
\end{align}
with the identifications $\kappa = t_1 t_2 t_3$, $t_4 = q/ \sqrt{\kappa}$, $t_5 = 1/(q \sqrt{\kappa})$; finally after symmetrization we set $q = - p_0 p_1 \cdots p_{n-1}$.
\end{conjecture}
We have checked this formula up to terms of degree five for $\complex^2 / \zed_n \times \complex$ with $n=2,3,4,5,6$.

\paragraph{Refined limit.}

Now we take a refined limit by sending $t_3 \longrightarrow 0$, $t_1 \longrightarrow 0$ in such a way that $t_3$ approaches zero much faster and $t_2 \longrightarrow \infty$ in such a way that the product $\kappa = t_1 t_2 t_3$ remains constant.

First recall that
\be
\ahat{w} \longrightarrow \left\{ \begin{matrix} - \kappa^{-1/2} & \text{if} \ w \longrightarrow \infty \\ 
 - \kappa^{1/2} & \text{if} \ w \longrightarrow 0 \end{matrix} \right.
\ee
Next we have the limits 
\begin{align}
\frac{t_1^{n-k}}{t_2^k} \longrightarrow 0 \, , \qquad \frac{t_2^{k+1}}{t_1^{n-1-k}} \longrightarrow \infty  \, ,
\end{align}
since $0 \le k \le n-1$. Therefore the contributions of each ratio to the refined limit exactly compensate each other for each $k$ and the leading term is always determined by the behaviour of $t_3$. 
Therefore each term in $\mathscr{F}_{\complex^2 / \zed_n \times \complex}$ contributes equally. If we set $t_4 = q/ \sqrt{ \kappa}$, $t_5 = 1/(q \sqrt{ \kappa})$, then 
\be
\mathscr{F}_{\complex^2 / \zed_n \times \complex} (t_1 , t_2, t_3 , t_4 , t_5) \longrightarrow n (- \kappa^{1/2}) \frac{1}{\kappa^{1/2}\, (1-q/ \sqrt{\kappa})(1-1/(q \sqrt{ \kappa}))} = n \, {\kappa}^{1/2} \, \frac{q}{(1-q \sqrt{\kappa}) (1 - q/\sqrt{\kappa})} 
\ee

In the function $\mathscr{F}_{\complex^2 / \zed_n \times \complex}^{(r)}$ the only term relevant to the refined limit is the $t_3$ dependent ratio. Overall this function becomes
\begin{align}
\mathscr{F}_{\complex^2 / \zed_n \times \complex}^{(r)} \left( \{ p_i \} , t_1 , t_2 , t_3 , t_4 , t_5 \right) \longrightarrow \sum_{0 < r \le s < n} \left( p_{[r,s]} + \frac{1}{p_{[r,s]}} \right) (\kappa^{1/2}) \frac{q}{(1-q \sqrt{\kappa}) (1 - q/\sqrt{\kappa})}
\end{align}

By writing all the pieces together we find the full refined limit
\be
Z_{\complex^2 / \zed_n \times \complex}^{ref} \left( \{ p_i \} ; \kappa \right) = \sfS^\bullet \left( 
\left( n + \sum_{0 < r \le s < n} \left( p_{[r,s]} + \frac{1}{p_{[r,s]}} \right) \right) \
 {\kappa}^{1/2} \, \frac{q}{(1-q \sqrt{\kappa}) (1 - q/\sqrt{\kappa})} 
\right)
\ee
where again after symmetrization we set $q = - p_0 p_1 \cdots p_{n-1}$.

We can now write this generating function in product formula. First note that
\begin{align} \label{genMac}
\sfS^\bullet x \, \frac{\kappa^{1/2} q}{(1-q \sqrt{\kappa}) (1 - q/\sqrt{\kappa})} 
 = \prod_{i,j=1}^\infty \left( 1 -x \,  q^{i+j-1} \, \sqrt{\kappa}^{j-i+1} \right)^{-1} =: M (x , q, \sqrt{\kappa})
\end{align}
where the generalized MacMahon function is defined in \eqref{RefinedM}; to ease the notation in this section we set $M_{\delta = 1} (x , q, \sqrt{\kappa})
=M (x , q, \sqrt{\kappa})
$ . Using repeatedly this identity we find
\begin{align}
Z_{\complex^2 / \zed_n \times \complex}^{ref} & \left( \{ p_i \} ; \kappa \right) =  M \left(1 , ( - p_0 p_1 \cdots p_{n-1}) , \sqrt{\kappa} \right)^n \times 
\cr & \times 
\prod_{0 < r \le s < n} M (p_{[r,s]} , ( - p_0 p_1 \cdots p_{n-1}), \sqrt{\kappa}) \ M ( \frac{1}{p_{[r,s]}} , ( - p_0 p_1 \cdots p_{n-1}), \sqrt{\kappa})
\end{align}

As in the conifold case, the results disagree with the literature \cite{MN} in the choice of refinement of the ``degree zero'' part, the factor $M \left(1 , ( - p_0 p_1 \cdots p_{n-1}) , \sqrt{\kappa} \right)^n$. However as already mentioned in \cite{MN} and discussed above, this has to be expected and is due to the non-compactness of the relevant moduli spaces. Most likely there exists a choice of the refined limit which reproduces the results of \cite{MN}, but we haven't managed to find it so far.

\subsection{The case $\complex^3 / \zed_2 \times \zed_2$}

\paragraph{Geometry and representation theory.}

In this case the orbifold acts as
\begin{align}
g_1 (z_1 , z_2 , z_3) &= (-z_1 , - z_2 , z_3) \cr
g_2 (z_1 , z_2 , z_3) &= (-z_1 , z_2 , -z_3) \cr
g_3 (z_1 , z_2 , z_3) &= (z_1 , - z_2 , - z_3)
\end{align}
and $g_0$ the identity. The crepant resolution given by $\mathrm{Hilb}_{\zed_2 \times \zed_2} (\complex^3, 4)$ has three compact curves and no compact divisors. The action has non trivial weights $r_1 = (1,1,0)$, $r_2=(1,0,1)$ and $r_3 = r_1 + r_2$. There are four corresponding irreducible representations $\rho_r$ and the tensor product decomposition $\rho_r \otimes Q$ with the defining representation gives 
\begin{equation}
\big(a_{rs}^{(1)}\big) = \left(  \begin{matrix}  0 & 1 & 1 & 1 \\ 1 &
    0 & 1 & 1 \\ 1 & 1 & 0 & 1 \\ 1 & 1 & 1 & 0 \end{matrix}\right) \ .
\label{ars1Z2Z2}\end{equation}
The resulting quiver $\sfQ$ is
\begin{equation}
\vspace{4pt}
\begin{xy}
\xymatrix@C=20mm{
& \ v_0 \ \bullet \ \ar@/^/[ddr] \ar@/_1pc/[ddl]  \ar@/^/[d]& \\
& \ v_3 \ \bullet \ \ar@/^/[u] \ar@/^/[dl] \ar@/^/[dr]& \\
v_1 \ \bullet \ \ar@/^/[uur] \ar@/^/[ur] \ar@//[rr] & & \ \bullet \ v_2  \ar@/_1pc/[uul] \ar@/^/[ul] \ar@/^/[ll]
}
\end{xy}
\vspace{4pt}
\end{equation}
The framed quiver $\hat{\sfQ}$ has an extra node connected by a single arrow to the node $v_0$. A tilting bundle for this geometry can be obtained from the formalism discussed in Section 1.1 of \cite{nagao}.

\paragraph{BPS partition function.}

The representation space decomposes as
\be
V_\pi = V_{0,0} + V_{0,1} + V_{1,0} + V_{1,1}
\ee
with
\be
V_{r_1,r_2} = \sum_{(i,j,k) \in \pi} t_1^{-i+1} t_2^{-j+1} t_3^{-k+1} \, \delta_{i+k=r_1 \!\!\!\! \mod 2}  \,  \delta_{j+k=r_2 \!\!\!\! \mod 2} 
\ee
As a consequence the virtual tangent space is
\begin{align}
(T_{\pi}^{\rm vir})^{\zed_2 \times \zed_2} = & V_{0,0} - V_{0,0} (t_1 \, t_2 \, t_3) - \Big[ (V_{0,0} \, \overline{V}_{0,0}+V_{1,0} \, \overline{V}_{1,0}+V_{0,1} \, \overline{V}_{0,1}+V_{1,1} \, \overline{V}_{1,1}) (1 - t_1 \, t_2 \, t_3)
\cr & - (V_{1,0} \, \overline{V}_{0,0}+V_{0,0} \, \overline{V}_{1,0}+V_{0,1} \, \overline{V}_{1,1}+V_{1,1} \, \overline{V}_{0,1}) (t_1 - t_2 \, t_3) 
\cr & - (V_{0,1} \, \overline{V}_{0,0}+V_{0,0} \, \overline{V}_{0,1}+V_{1,0} \, \overline{V}_{1,1}+V_{1,1} \, \overline{V}_{1,0}) (t_2 - t_1 \, t_3)
\cr &  - (V_{0,0} \, \overline{V}_{1,1}+V_{1,1} \, \overline{V}_{0,0}+V_{1,0} \, \overline{V}_{0,1}+V_{0,1} \, \overline{V}_{1,0}) (t_3 - t_1 \, t_2) \Big] \, .
\end{align}
From the virtual tangent space and the classification of fixed points we can compute the partition function, defined as
\be
\sfZ_{\complex^3 / \zed_2 \times \zed_2} \left( \{ p_i \} , \{ t_i \} \right) = \sum_{n} \sum_{|\pi| = n} p_{0}^{|\pi_{00}|} \, p_1^{|\pi_{10}|} \, p_2^{|\pi_{01}|} \, p_3^{|\pi_{11}|} \, \hat{\mathsf{a}} \left( (T_\pi^{\mathrm{vir}})^{\zed_2 \times \zed_2} \right) \, .
\ee
Explicitly the first few terms read
\begin{align}
\sfZ_{\complex^3 / \zed_2 \times \zed_2} & \left( \{ p_i \} , \{ t_i \} \right)= 
1+p_0+
\frac{p_0 p_1 \left(\kappa-t_1^2\right)}{\sqrt{\kappa}
   \left(-1+t_1^2\right)} +\frac{p_0
   p_2 \left(\kappa-t_2^2\right)}{\sqrt{\kappa}
   \left(-1+t_2^2\right)}+\frac{p_0
   p_3 \left(\kappa-t_3^2\right)}{\sqrt{\kappa}
   \left(-1+t_3^2\right)}
+ p_0 p_1 p_2 \cr & +p_0 p_1 p_3+p_0
   p_2 p_3  +\frac{p_0^2 p_1
   \left(\kappa-t_1^2\right)}{\sqrt{\kappa} \left(-1+t_1^2\right)} 
     +\frac{p_0^2 p_2
   \left(\kappa-t_2^2\right)}{\sqrt{\kappa} \left(-1+t_2^2\right)}+\frac{p_0^2 p_3
   \left(\kappa-t_3^2\right)}{\sqrt{\kappa} \left(-1+t_3^2\right)} + \cdots
\end{align}

\paragraph{Symmetrized form.}

Now we write down the orbifold function, whose symmetrization reproduces the partition function. Consider first the function
\begin{align}
\mathscr{F}_{\complex^3 / \zed_2 \times \zed_2} \left[ t_1 , t_2 , t_3 , t_4, t_5 \right] & = 
\mathscr{F}_{\complex^3} \left[ t_1^2 , t_2^2  , \frac{t_3}{t_1 t_2} ,  t_4 , t_5 \right]
+\mathscr{F}_{\complex^3} \left[  t_1^2, \frac{t_2}{t_1 t_3} ,  t_3^2 ,  t_4 , t_5 \right]
\cr & +\mathscr{F}_{\complex^3} \left[ \frac{t_1}{t_2 t_3} , t_2^2  , t_3^2 ,  t_4 , t_5 \right]
+\mathscr{F}_{\complex^3} \left[ \frac{t_1 t_2}{t_3} , \frac{t_1 t_3}{t_2}  , \frac{t_2 t_3}{t_1}  ,  t_4 , t_5 \right]
\end{align}
Such a function is related to the decomposition
\begin{align}
\frac14 \sum_{i=0}^3 & \frac{1}{(1- g_i t_1) (1-g_i t_2) (1-g_i t_3)} = \frac{1}{\left(1-t_1^2\right) \left(1-t_2^2\right)
   \left(1-\frac{t_3}{t_1 t_2}\right)}+\frac{1}{\left(1-t_1^2\right) \left(1-t_3^2\right)
   \left(1-\frac{t_2}{t_1
   t_3}\right)}
   \cr & +\frac{1}{\left(1-t_2^2\right) \left(1-t_3^2\right)
   \left(1-\frac{t_1}{t_2 t_3}\right)}+\frac{1}{\left(1-\frac{t_1
   t_2}{t_3}\right) \left(1-\frac{t_1 t_3}{t_2}\right)
   \left(1-\frac{t_2 t_3}{t_1}\right)}
\end{align}
which generates invariant monomials. Secondly define
\begin{align}
& \mathscr{F}_{\complex^3 / \zed_2 \times \zed_2}^{(r)} \left[ \{ p_i \} , t_1 , t_2 , t_3 , t_4 , t_5 \right] = 
\frac{1}{\left(\sqrt{t_4}-\frac{1}{\sqrt{t_4}}\right)
   \left(\sqrt{t_5}-\frac{1}{\sqrt{t_5}}\right)} 
   \cr & \times \Big[
   \frac{\left(\frac{\sqrt{\kappa}}{t_3}-\frac{t_3}{\sqrt{\kappa}}\right)
   \left(p_1 p_2
   +\frac{1}{p_1
   p_2}\right)}{t_3-\frac{1}{t_3}}
   +\frac{\left(\frac{\sqrt{\kappa}}{t_2}-\frac{t_2}{\sqrt{\kappa}}\right) \left(p_1 p_3+\frac{1}{p_1
   p_3}\right)}{t_2-\frac{1}{t_2}}+\frac{\left(\frac{\sqrt{\kappa}}{t_1
   }-\frac{t_1}{\sqrt{\kappa}}\right) \left(p_2 p_3+\frac{1}{p_2
   p_3}\right)}{t_1-\frac{1}{t_1}}
   \cr & +p_1 p_2
   p_3+\frac{1}{p_1 p_2
   p_3}+p_1+\frac{1}{p_1}+p_2+\frac{1}{p_2}+p_3+\frac
   {1}{p_3}
   \Big] \, .
\end{align}
When comparing to the large radius result we expect that the parameters $p_1$, $p_2$ and $p_3$ correspond to the three compact curves in the resolved geometry.

We conjecture that
\begin{align}
\sfZ_{\complex^3 / \zed_2 \times \zed_2} \left( \{ p_i \} , \{ t_i \} \right)  = \sfS^\bullet \left( 
\mathscr{F}_{\complex^3 / \zed_2 \times \zed_2} \left[ t_1 , t_2 , t_3 , t_4, t_5 \right] +  \mathscr{F}_{\complex^3 / \zed_2 \times \zed_2}^{(r)} \left[ \{ p_i \} , t_1 , t_2 , t_3 , t_4 , t_5 \right]
\right)
\end{align}
with the usual identifications $\kappa = t_1 t_2 t_3$, $t_4 = q/ \sqrt{ \kappa}$, $t_5 = 1/(q \sqrt{ \kappa})$ and, after symmetrization, $q = - p_0 p_1 p_2 p_3$.

\paragraph{Refined limit.}

By scaling away the toric weights with the same procedure of A-fibered singularities we find
\begin{align}
Z_{\complex^3 / \zed_2 \times \zed_2}^{ref} \left( \{ p_i \} ,\kappa \right)  & = \sfS^\bullet \Big[ \frac{q \, \kappa^{1/2}}{(1-\frac{q}{\kappa^{1/2}})(1-\kappa^{1/2} \, q)} \Big( 3 + \frac{1}{\kappa}+ \left( p_1 \, p_2 + \frac{1}{p_1 \, p_2} \right) + \left( p_2 \, p_3 + \frac{1}{p_2 \, p_3} \right) 
\cr 
& + \frac{1}{\kappa} \left( p_1 \, p_3 + \frac{1}{p_1 \, p_3} \right) - \frac{1}{\kappa^{1/2}} \left( p_1 + \frac{1}{p_1} \right)  - \frac{1}{\kappa^{1/2}} \left( p_2 + \frac{1}{p_2} \right)  - \frac{1}{\kappa^{1/2}} \left( p_3 + \frac{1}{p_3} \right) \cr & -\frac{1}{\kappa^{1/2}} \left( p_1 p_2 p_3 + \frac{1}{p_1 p_2 p_3} \right) \Big) \Big]
\end{align}
where we set $q = -p_0 \, p_1 \, p_2 \, p_3$ after symmetrization. Proceeding as before, we can also write down a product form for the partition function
\begin{align}
Z_{\complex^3 / \zed_2 \times \zed_2}^{ref} \left( \{ p_i \} ,\kappa \right)= & M (1 , q, \sqrt{\kappa})^3 \ M (1/\kappa , q, \sqrt{\kappa}) \ M (p_1 \, p_2 , q, \sqrt{\kappa}) \ M (\frac{1}{p_1 \, p_2} , q, \sqrt{\kappa}) 
\cr & M (p_2 \, p_3 , q, \sqrt{\kappa}) \ M (\frac{1}{p_2 \, p_3} , q, \sqrt{\kappa}) \ M (\frac{1}{\kappa} p_1 \, p_3 , q, \sqrt{\kappa}) \ M (\frac{1}{\kappa} \frac{1}{p_1 \, p_3} , q, \sqrt{\kappa}) 
\cr & 
M (\frac{1}{\sqrt{\kappa}} \, p_1 , q, \sqrt{\kappa})^{-1} \ M (\frac{1}{\sqrt{\kappa}} \, p_2 , q, \sqrt{\kappa})^{-1} \ M (\frac{1}{\sqrt{\kappa}} \, p_3 , q, \sqrt{\kappa})^{-1}
\cr &
M (\frac{1}{\sqrt{\kappa}} \, \frac{1}{p_1} , q, \sqrt{\kappa})^{-1} \ M (\frac{1}{\sqrt{\kappa}} \, \frac{1}{p_2} , q, \sqrt{\kappa})^{-1} \ M (\frac{1}{\sqrt{\kappa}} \, \frac{1}{p_3} , q, \sqrt{\kappa})^{-1} \, 
\cr & 
M (\frac{1}{\sqrt{\kappa}} \, p_1 \, p_2 \, p_3 , q, \sqrt{\kappa})^{-1} \ M (\frac{1}{\sqrt{\kappa}} \, \frac{1}{p_1 \, p_2 \, p_3} , q, \sqrt{\kappa})^{-1}  \, .
\end{align}
where the generalized MacMahon function was introduced in \eqref{genMac}.

\subsection{The $\complex^3 / \zed_3$ orbifold}

Consider the orbifold $\mathbb{C}^3 / \mathbb{Z}_3$. In this case no closed formula is known even in the case of the (unrefined) topological string, and therefore we do not expect one in the K-theory setup.

\paragraph{Geometry and representation theory.}

We let the generator $g$ of $\mathbb{Z}_3$ act on $\mathbb{C}^3$ as $g (z_1 , z_2 , z_3) = ( \zeta z_1,\zeta z_2,\zeta z_3)$ with $\zeta^3=1$. The resulting orbifold $\complex^3 / \zed_3$ has a geometric crepant resolution given by the total space of $\cO_{\PP^2} (-3) \longrightarrow \PP^2$. This geometry contains a $\PP^2$ as a compact divisor, and three torus invariant rational curves.

From the representation theory data one can construct the quiver
\begin{equation}
\vspace{4pt}
\begin{xy}
\xymatrix@C=8mm{
& \ v_0 \ \bullet \ \ar@/^/[ddl] \ar@/_0.5pc/[ddl] \ar@//[ddl]  & \\
& & \\
v_1 \ \bullet \ \ar@//[rr] \ar@/^/[rr]  \ar@/_/[rr]   & &  \ \bullet \ v_2  \ar@/^/[uul] \ar@/_0.5pc/[uul] \ar@//[uul] 
}
\end{xy}
\vspace{4pt}
\label{quiverC3Z3}\end{equation}
Indeed in this case $Q = \rho_1 \oplus \rho_1 \oplus \rho_1$, from which we see from (\ref{tensordecomp}) that
\begin{equation}
 a_{ab}^{(1)}= \left( \begin{matrix} 0 & 0 & 3 \\ 3 & 0 & 0 \\  0 & 3 & 0 \end{matrix} \right)
\label{C3Z3matrices}\end{equation}
The framing consists in an extra node connected with a single arrow to $v_0$.

\paragraph{BPS partition function.} The representation space has the decomposition $V = V_0 + V_1 + V_2$, where
\be
V_r =   \sum_{(i,j,k) \in \pi} t_1^{-i+1} t_2^{-j+1} t_3^{-k+1} \delta_{i+j+k=r \!\!\!\! \mod 3}
\ee
The virtual tangent space is given by
\begin{align}
(T_{\pi}^{\rm vir})^{\Gamma} = 
& V_0 - \overline{V}_0 (t_1 \, t_2 \, t_3) - \Big[ (V_0 \, \overline{V}_0 + V_1 \, \overline{V}_1 + V_2 \, \overline{V}_2 ) (1 - t_1 \, t_2 \, t_3) 
\cr & + (V_0 \, \overline{V}_2 + V_1 \, \overline{V}_0 + V_2 \, \overline{V}_1) (-t_1 - t_2 - t_3) 
\cr & + (V_0 \, \overline{V}_1 + V_1 \, \overline{V}_2 + V_2 \, \overline{V}_0) (t_1 \, t_2 + t_1 \, t_3 + t_2 \, t_3) \Big] \, .
\end{align}
The partition function is now
\be
\sfZ_{\complex^3 / \zed_3} \left( \{ p_i \} , \{ t_i \} \right)  = \sum_{n} \sum_{|\pi| = n} p_{0}^{|\pi_{0}|} \, p_1^{|\pi_{1}|} \, p_2^{|\pi_{2}|} \, \hat{\mathsf{a}} \left( (T_\pi^{\mathrm{vir}})^{\Gamma} \right) \, .
\ee
Explicitly the few first terms are
\begin{align}
\sfZ_{\complex^3 / \zed_3} & \left( \{ p_i \} , \{ t_i \} \right) = 
1+p_0+\frac{p_0 p_1 (\kappa t_1-t_2) (\kappa t_1-t_3)}{\kappa
   (t_1-t_2) (t_1-t_3)} +\frac{p_0 p_1 (t_1-\kappa t_2) (\kappa
   t_2-t_3)}{\kappa (t_1-t_2) (t_2-t_3)}
   \\ & +\frac{p_0 p_1 (t_1-\kappa t_3) (-t_2+\kappa
   t_3)}{\kappa (t_1-t_3) (-t_2+t_3)} 
     -\frac{p_0 p_1 p_2
   \left(-\kappa+t_1^3\right) (\kappa t_1-t_2) (\kappa t_1-t_3)}{\kappa^{3/2}
   \left(-1+t_1^3\right) (t_1-t_2)
   (t_1-t_3)}
   \cr & -\frac{p_0
   p_1 p_2 (t_1-\kappa t_2) \left(-\kappa+t_2^3\right) (\kappa
   t_2-t_3)}{\kappa^{3/2} (t_1-t_2) \left(-1+t_2^3\right)
   (t_2-t_3)} -\frac{p_0 p_1
   p_2 (t_1-\kappa t_3) (-t_2+\kappa t_3)
   \left(-\kappa+t_3^3\right)}{\kappa^{3/2} (t_1-t_3) (-t_2+t_3)
   \left(-1+t_3^3\right)}
   \cr & +\frac{p_0 p_1^2 (\kappa t_1-t_3) (-\kappa
   t_2+t_3)}{\kappa (t_1-t_3) (-t_2+t_3)}+\frac{p_0
   p_1^2 (t_1-\kappa t_2) (t_1-\kappa t_3)}{\kappa (t_1-t_2)
   (t_1-t_3)}+\frac{p_0 p_1^2 (\kappa t_1-t_2)
   (t_2-\kappa t_3)}{\kappa (t_1-t_2)
   (t_2-t_3)}
   + \cdots \nonumber
\end{align}

\section{Discussion}

We conclude with some comments and some open problems.

\subsection{Towards M2 branes on NCCRs}

The results of the previous sections should admit an interpretation as the theory of M2 branes propagating on noncommutative crepant resolutions. Indeed the main goal of the M2-brane index computations is to provide some mathematical understanding of certain degrees of freedom of M-theory. While it is not yet clear how precisely to attain this goal, we can still try to make some progress in understanding the noncommutative theory by transferring large radius results to the singular non-geometric limit. The following discussion will be necessarily somewhat vague.

Roughly speaking the proposal made in \cite{NO} for the membrane moduli space consists of the moduli space of maps $f \, : C \longrightarrow Z$ from 1-dimensional Cohen-Macaulay schemes into the smooth Calabi-Yau fivefold, with certain stability conditions that ensure that the Euler characteristic $\chi (\cO_C)$ is bounded both from above and from below. More precisely the moduli space $\mathsf{M2} (Z)$ parametrizes such maps together with a certain subsheaf $N$ of the normal sheaf $N_f$. Such moduli space is endowed with a symmetrized virtual structure sheaf $\cO_{\mathsf{M2}}$. Conjecturally virtual counts over this moduli space should reproduce the K-theoretical Donaldson-Thomas theory of $Z$.

Our results should point the way to define the analogous object in the noncommutative regime $\mathsf{NCM2} (Z)$, which describe membranes degrees of freedom when $Z$ develops a singularity. This means that it should be possible to interpret our results as a theory of membranes in a noncommutative geometry, possibly obtained as a phase of a membrane worldvolume theory. One could expect that such a theory is related to maps from a curve $C$ into the noncommutative algebra $\mathsf{A}$ which resolves the singularity, where the membrane is wrapping the three manifold $C \times S^1$. The situation is somewhat analogous to what happens in the study of three dimensional gauge theories, as discussed for example in \cite{Okounkov}. If one considers a supersymmetric gauge theory on $C \times S^1$ then under certain assumptions the Higgs branch has the structure of a Nakajima variety. At low energies the theory reduces to the K-theory of quasimaps $f : C \dashrightarrow \rm{Nak}$, defined as sections of certain bundles $\cV$, $\cW$ on $C$ which obey certain moment map relations $\mu = 0$. Here $\cV_i = f^* V_i$ where $V_i$ are the vector spaces which enter in the definition of the Nakajima quiver variety. In our case the relevant target space is not a Nakajima variety but we can still consider the moduli space of representations of the relevant quiver. Indeed this plays the same role in six-dimensional instanton counting that Nakajima varieties play in four dimensional instanton counting \cite{Cirafici:2012qc}.

Roughly speaking one would study collections of vector bundles $\cV_i$ on $C$ of fixed rank, together with a section $f \in H^0 (C , \mathcal{M}^\Gamma)$ which obeys $\mu = 0$, with
\be
\cM^\Gamma = 
\mathrm{Hom}_{\Gamma} (\cV , \cQ \otimes \cV) \ 
 \oplus \ \mathrm{Hom}_{\Gamma} (\cV , \mbox{$\bigwedge^3$} \cQ \otimes \cV) \ 
\oplus \ \mathrm{Hom}_{\Gamma} (\cW , \cV) \ 
\ee
where the product of toric weights is interpreted as $\hbar$. The moment map $\mu=0$ is the one obtained by the F-term equations derived from the superpotential  \cite{Cirafici:2010bd}.

The associated enumerative problem appears to be interesting and somewhat natural. However there are two serious drawbacks: to begin with it is not clear at all if it is mathematically well posed; secondly the physical motivations are based on an analogy, but this is rather thin and there is no reason that the worldvolume theory of the membrane will have a phase where the moduli space of vacua is described by quasimaps.

Another option would be to transfer directly the construction of \cite{NO} from large radius to the singular phase. In the case of orbifolds we can use the McKay correspondence, or more generically the fact that the derived category  $\rm D^b (coh X)$ of the smooth threefold is derived equivalent to the derived category of modules over the path algebra of the quiver $\rm D^b (\sfA-mod)$. This  equivalence is given by a tilting sheaf and under favourable conditions it descends at the level of K-theory. For example for the resolved conifold $Y$ the tilting generator is $T = \cO_Y \oplus \pi^* \cO_{P^1} (1)$. Then $\mathbb{R} \Hom (T , -) \, : \, \rm D^b (coh (Y)) \longrightarrow D^b (\mathsf{A}-mod)$ induces the derived equivalence, and $\sfA = \End_Y (T)$. The tilting generator has the form $T = \mathsf{P}_\circ \oplus \mathsf{P}_{\bullet}$ where $P_i = \sfA \, \mathsf{e}_i$ are the projective modules and the $\mathsf{e}_i$ are the idempotents. 

The derived equivalence descends at the level of K-theory to the isomorphism $\rm [RF (-)] \, : \, K (coh Y) \longrightarrow K (\sfA-mod)$ (since this discussion is rather qualitative we ignore when opposite categories appear), given by $[RF (-)] = \sum_i (-1)^i \, \Ext^i_Y (T , -)$. Quasi-isomorphic complexes map to the same K-theory class. In the large radius limit we are looking at maps  $f$ from schemes $C$ to $Z$. We can replace the curve with the sheaf $f_* \cO_C$, and then via the derived equivalence map the two sheaves $f_* \cO_C$ and $\mathcal{N}$ into complexes of $\sfA$-modules, with K-theory classes  $\rm [RF (\cO_C)]$ and  $\rm [RF (\mathcal{N})]$.


In the case of orbifolds the above procedure is essentially a generalization of the McKay correspondence, reviewed in \cite{Cirafici:2012qc} in a similar context. At the level of K-theory this maps complexes of vector bundles which are exact outside the exceptional locus to $\Gamma$-equivariant sheaves supported at the origin of $\complex^3$. Therefore this discussion predicts that the moduli space $\mathsf{NCM2}$ of M2 branes on the noncommutative geometry can be understood as the $\Gamma$-fixed locus $\mathsf{M2}_\Gamma (\complex^5)$ of the flat space moduli space $\mathsf{M2} (\complex^5)$. Note that while the discussion above is rather vague, this proposal is concrete. In particular it should be possible to analyse this moduli space concretely to check the above quasi-map proposal.

We hope to return to this problem in the near future
\subsection{Some open problems}

We summarize a few more concrete open problems that are currently under investigation

\begin{itemize}
\item Perhaps the most interesting problem is a systematic study of the wall-crossing behaviour of the M2 index. In this paper we have computed the index in the noncommutative crepant resolution chamber. We believe our techniques can be extended to other chambers, especially those which arise from this chamber via quiver mutations. Work is in progress in this direction and we hope to report about it soon \cite{MWC}.
\item Ordinary Donaldson-Thomas theory associated to quivers can also be rephrased in terms of a dimer model, a certain statistical mechanical model associated with the quiver \cite{Ooguri:2009ijd}. It should be possible to generalize this model to the case where the weight of a dimer configuration is given by the localized A-roof genus of the configuration, computed at a fixed point in the associated quiver moduli space. It would be interesting to understand if such a model has hidden algebraic structures, perhaps related to those uncovered in \cite{Li:2020rij}.
\item Donaldson-Thomas theory, from a gauge theory point of view, can be generalized to include defects \cite{Cirafici:2013nja}. Roughly speaking the idea is to specify singularities for the gauge connection over a divisor. Such singularities can be classified in term of parabolic sheaves and the Donaldson-Thomas problem is reduced to counting ideal sheaves with a prescribed parabolic structure. Also this problem can be rephrased in terms of representations of a certain quiver, and our techniques certainly apply to that case. 
\item In the case of orbifolds the higher rank Donaldson-Thomas invariants admit a generalization as Donaldson-Thomas invariants of type $\mathbf{w}$, where $\mathbf{w}$ is the vector which specifies the rank of the D6 brane theory as well as the boundary conditions at infinity \cite{Cirafici:2010bd}. It would be interesting to study the K-theoretic generalization to this setting. In particular it is natural to propose that the same vector $\mathbf{w}$ specifies boundary conditions at infinity for M2 branes. 
\end{itemize}

\section*{Acknowledgements}

I am grateful to Vasily Pestun for many discussions and collaboration in the early stages of this work. I also thank Sergej Monavari for many comments and catching a few typos. I am a member of INDAM-GNFM and I am supported by INFN via the Iniziativa Specifica GAST.  The author also acknowledges the support of IH\'ES during a visit.
The research of M.C. on this project has received funding from the European Research Council
(ERC) under the European Union's Horizon 2020 research and innovation programme (QUASIFT
grant agreement 677368).

\section*{Data Availability Statement}

All data generated or analysed during this study are included in this published article (and its supplementary information files). Alternatively they are freely available for download at \url{https://cirafici.dmg.units.it/MembranesDT.zip}, as well as with the arxiv submission at \url{https://arxiv.org/abs/2111.01102}.

\section*{Conflict of Interest Statement}

The corresponding author states that there is no conflict of interest.

\end{document}

%% file: conifoldFP.pdf_tex
\begingroup%
  \makeatletter%
  \providecommand\color[2][]{%
    \errmessage{(Inkscape) Color is used for the text in Inkscape, but the package 'color.sty' is not loaded}%
    \renewcommand\color[2][]{}%
  }%
  \providecommand\transparent[1]{%
    \errmessage{(Inkscape) Transparency is used (non-zero) for the text in Inkscape, but the package 'transparent.sty' is not loaded}%
    \renewcommand\transparent[1]{}%
  }%
  \providecommand\rotatebox[2]{#2}%
  \newcommand*\fsize{\dimexpr\f@size pt\relax}%
  \newcommand*\lineheight[1]{\fontsize{\fsize}{#1\fsize}\selectfont}%
  \ifx\svgwidth\undefined%
    \setlength{\unitlength}{789.27163696bp}%
    \ifx\svgscale\undefined%
      \relax%
    \else%
      \setlength{\unitlength}{\unitlength * \real{\svgscale}}%
    \fi%
  \else%
    \setlength{\unitlength}{\svgwidth}%
  \fi%
  \global\let\svgwidth\undefined%
  \global\let\svgscale\undefined%
  \makeatother%
  \begin{picture}(1,0.70707072)%
    \lineheight{1}%
    \setlength\tabcolsep{0pt}%
    \put(0,0){\includegraphics[width=\unitlength,page=1]{conifoldFP.pdf}}%
    \put(0.55932547,0.65293635){\color[rgb]{0,0,0}\makebox(0,0)[lt]{\lineheight{0}\smash{\begin{tabular}[t]{l}$t_{a_1}$\end{tabular}}}}%
    \put(0.00190862,0.37494071){\color[rgb]{0,0,0}\makebox(0,0)[lt]{\lineheight{0}\smash{\begin{tabular}[t]{l}$t_{b_2}$\end{tabular}}}}%
    \put(0.7375278,0.38349447){\color[rgb]{0,0,0}\makebox(0,0)[lt]{\lineheight{0}\smash{\begin{tabular}[t]{l}$t_{b_1}$\end{tabular}}}}%
    \put(0.50230073,0.01425922){\color[rgb]{0,0,0}\makebox(0,0)[lt]{\lineheight{0}\smash{\begin{tabular}[t]{l}$t_{a_2}$\end{tabular}}}}%
  \end{picture}%
\endgroup%

%% file: membranesquivers11.bbl
\begin{thebibliography}{99}

\bibitem{kai}
K.~Behrend, ``Donaldson-Thomas type invariants via microlocal geometry", Annals of Mathematics, 2nd Ser., \textbf{170} (3): 1307-1338 [arXiv:math/0507523].

\bibitem{fantechi}
K.~Behrend and B.~Fantechi,
``Symmetric obstruction theories and Hilbert schemes of points
on threefolds," Algebra Number Theory 2 (2008) 313-345 [arXiv:math.AG/0512556].

\bibitem{Benini:2018hjy}
F.~Benini, G.~Bonelli, M.~Poggi and A.~Tanzini,
``Elliptic non-Abelian Donaldson-Thomas invariants of $\mathbb{C}^3$,''
JHEP \textbf{07} (2019), 068
[arXiv:1807.08482 [hep-th]].

\bibitem{Bonelli:2020gku}
G.~Bonelli, N.~Fasola, A.~Tanzini and Y.~Zenkevich,
``ADHM in 8d, coloured solid partitions and Donaldson-Thomas invariants on orbifolds,''
[arXiv:2011.02366 [hep-th]].

\bibitem{Cirafici:2018jor}
M.~Cirafici,
``On Framed Quivers, BPS Invariants and Defects,''
Confluentes Mathematici 9 (2017) 2, 71-99 [arXiv:1801.03778 [hep-th]].

\bibitem{Cirafici:2019otj}
M.~Cirafici,
``Quantum Line Defects and Refined BPS Spectra,''
Lett. Math. Phys. \textbf{110} (2019) no.3, 501-531
[arXiv:1902.08586 [hep-th]].



\bibitem{Cirafici:2008sn}
  M.~Cirafici, A.~Sinkovics and R.~J.~Szabo,
  ``Cohomological gauge theory, quiver matrix models and Donaldson-Thomas theory,''
  Nucl.\ Phys.\ B {\bf 809} (2009) 452
  [arXiv:0803.4188 [hep-th]].

\bibitem{Cirafici:2010bd}
  M.~Cirafici, A.~Sinkovics and R.~J.~Szabo,
  ``Instantons, Quivers and Noncommutative Donaldson-Thomas Theory,''
  Nucl.\ Phys.\ B {\bf 853} (2011) 508
  [arXiv:1012.2725 [hep-th]].
  
\bibitem{Cirafici:2011cd}
  M.~Cirafici, A.~Sinkovics and R.~J.~Szabo,
  ``Instanton counting and wall-crossing for orbifold quivers,''
  Annales Henri Poincare {\bf 14} (2013) 1001
  [arXiv:1108.3922 [hep-th]].
%

\bibitem{Cirafici:2012qc}
M.~Cirafici and R.~J.~Szabo,
``Curve counting, instantons and McKay correspondences,''
J. Geom. Phys. \textbf{72} (2013), 54-109
[arXiv:1209.1486 [hep-th]].

\bibitem{Cirafici:2013nja}
M.~Cirafici,
``Defects in cohomological gauge theory and Donaldson-Thomas invariants,''
Adv. Theor. Math. Phys. \textbf{20} (2016), 945-1006
[arXiv:1302.7297 [hep-th]].

\bibitem{MWC}
M.~Cirafici. ``Wall-crossing and membranes'' \textit{to appear}.

\bibitem{mathe}
M.~Cirafici, supporting \textsc{mathematica} files available at \url{https://cirafici.dmg.units.it/MembranesDT.zip}, or as an ancillary files in the arXiv
submission.

\bibitem{DelZotto:2021gzy}
M.~Del Zotto, N.~Nekrasov, N.~Piazzalunga and M.~Zabzine,
``Playing with the index of M-theory,''
[arXiv:2103.10271 [hep-th]].

\bibitem{Descombes:2021snc}
P.~Descombes,
``Motivic DT invariants from localization,''
[arXiv:2106.02518 [math.AG]].

\bibitem{Chuang:2013wt}
W.~y.~Chuang, D.~E.~Diaconescu, J.~Manschot, G.~W.~Moore and Y.~Soibelman,
``Geometric engineering of (framed) BPS states,''
Adv. Theor. Math. Phys. \textbf{18} (2014) no.5, 1063-1231
[arXiv:1301.3065 [hep-th]].

\bibitem{DG}
  T.~Dimofte and S.~Gukov,
  ``Refined, Motivic, and Quantum,''
  Lett.\ Math.\ Phys.\  {\bf 91} (2010) 1
  [arXiv:0904.1420 [hep-th]].

\bibitem{thomas}
S. K. Donaldson and R. P. Thomas, ``Gauge theory in higher dimensions'', in \textit{The Geometric Universe; Science, Geometry, And The Work Of Roger Penrose}, Oxford University Press, 1998

\bibitem{Fasola:2020hqa}
N.~Fasola, S.~Monavari and A.~T.~Ricolfi,
``Higher rank K-theoretic Donaldson-Thomas theory of points,''
Forum Math. Sigma \textbf{9} (2021), e15
[arXiv:2003.13565 [math.AG]].

\bibitem{Iqbal:2003ds}
A.~Iqbal, N.~Nekrasov, A.~Okounkov and C.~Vafa,
``Quantum foam and topological strings,''
JHEP \textbf{04} (2008), 011
[arXiv:hep-th/0312022 [hep-th]].

\bibitem{Jafferis:2008uf}
D.~L.~Jafferis and G.~W.~Moore,
``Wall crossing in local Calabi Yau manifolds,''
[arXiv:0810.4909 [hep-th]].

\bibitem{Kononov:2019fni}
Y.~Kononov, A.~Okounkov and A.~Osinenko,
``The 2-Leg Vertex in K-theoretic DT Theory,''
Commun. Math. Phys. \textbf{382} (2021) no.3, 1579-1599
[arXiv:1905.01523 [math-ph]].

\bibitem{KS}
  M.~Kontsevich and Y.~Soibelman,
  ``Stability structures, motivic Donaldson-Thomas invariants and cluster transformations,''
  arXiv:0811.2435 [math.AG].

\bibitem{Li:2020rij}
W.~Li and M.~Yamazaki,
``Quiver Yangian from Crystal Melting,''
JHEP \textbf{11} (2020), 035
[arXiv:2003.08909 [hep-th]].

\bibitem{Maulik:2003rzb}
D.~Maulik, N.~Nekrasov, A.~Okounkov and R.~Pandharipande,
``Gromov\textendash{}Witten theory and Donaldson\textendash{}Thomas theory, I,''
Compos. Math. \textbf{142} (2006) no.05, 1263-1285
[arXiv:math/0312059 [math.AG]].

\bibitem{Maulik:2008nu}
D.~Maulik, A.~Oblomkov, A.~Okounkov and R.~Pandharipande,
``Gromov-Witten/Donaldson-Thomas correspondence for toric 3-folds,''  Invent. math. \textbf{186}, 435-479 (2011).
[arXiv:0809.3976 [math.AG]].

\bibitem{MMNS}
  A.~Morrison, S.~Mozgovoy, K.~Nagao and B.~Szendroi,
  ``Motivic Donaldson-Thomas invariants of the conifold and the refined topological vertex,'' Advances in Mathematics 230 (4-6) (2012)
  [arXiv:1107.5017 [math.AG]].

\bibitem{MN}
A.~Morrison and K.~Nagao
``Motivic Donaldson-Thomas invariants of toric small crepant resolutions,''
Algebra Number Theory 9 (2015) 767-813
arXiv:1110.5976 [math.AG]

\bibitem{Mozgovoy:2020has}
S.~Mozgovoy and B.~Pioline,
``Attractor invariants, brane tilings and crystals,''
[arXiv:2012.14358 [hep-th]].


\bibitem{reineke}
S.~Mozgovoy and M.~Reineke,
``On the noncommutative Donaldson-Thomas invariants arising from brane tilings,'' Adv. Math. {\bf 223} (2010) 1521 
[arXiv:0809.0117 [math.AG]].

\bibitem{nagao}
K.~Nagao, ``Derived categories of small toric Calabi-Yau 3-folds and counting invariants'',
The Quarterly Journal of Mathematics, Volume 63, Issue 4, December 2012, pages 965-1007 2012, [arXiv:0809.2994].

\bibitem{Nagao:2010kx}
K.~Nagao and H.~Nakajima,
``Counting invariant of perverse coherent sheaves and its wall-crossing,'' International Mathematics Research Notices, 17 (2011)
[arXiv:0809.2992 [math.AG]].

\bibitem{Nekrasov:2004vv}
N.~Nekrasov,
``A la recherche de la M-theorie perdue Z theory: Chasing M / f theory,'' In: Annual International Conference on Strings, Theory and Applications (Strings 2004) Paris, France, 28 June-July 2, 2004 (2004).
[arXiv:hep-th/0412021 [hep-th]].

\bibitem{Nekrasov:2017cih}
N.~Nekrasov,
``Magnificent four,''
Adv. Theor. Math. Phys. \textbf{24} (2020) no.5, 1171-1202
[arXiv:1712.08128 [hep-th]].

\bibitem{NO}
  N.~Nekrasov and A.~Okounkov,
  ``Membranes and Sheaves,'' Algebraic Geometry (2016) Vol. 3. No. 3. P. 320-369
  [arXiv:1404.2323 [math.AG]].

\bibitem{Nekrasov:2018xsb}
N.~Nekrasov and N.~Piazzalunga,
``Magnificent Four with Colors,''
Commun. Math. Phys. \textbf{372} (2019) no.2, 573-597
[arXiv:1808.05206 [hep-th]].

\bibitem{Okounkov}
  A.~Okounkov,
  ``Lectures on K-theoretic computations in enumerative geometry'' , in: Bezrukavnikov, Roman et al.
(Eds.), Geometry of Moduli Spaces and Representation Theory,
AMS and IAS, (2017), 
  arXiv:1512.07363 [math.AG].

\bibitem{Ooguri:2009ijd}
H.~Ooguri and M.~Yamazaki,
``Crystal Melting and Toric Calabi-Yau Manifolds,''
Commun. Math. Phys. \textbf{292} (2009), 179-199
[arXiv:0811.2801 [hep-th]].

\bibitem{Pomoni:2021hkn}
E.~Pomoni, W.~Yan and X.~Zhang,
``Tetrahedron instantons,''
[arXiv:2106.11611 [hep-th]].

\bibitem{szendroi}
  B.~Szendroi,
  ``Non-commutative Donaldson-Thomas theory and the conifold,''
  Geom.\ Topol.\  {\bf 12} (2008) 1171
  [arXiv:0705.3419 [math.AG]].

\bibitem{vandenbergh}
M.Van den Bergh ``Non-commutative crepant resolutions'', The legacy of Niels Henrik Abel, 749-770, Springer, Berlin, 2004.

\end{thebibliography}
